\documentclass[11pt]{article}

\usepackage[iso-8859-7]{inputenc}
\usepackage{epsfig}
\usepackage{graphicx}
\usepackage{epstopdf}
\usepackage{amsfonts}
\usepackage{amssymb}
\usepackage{amsthm}
\usepackage{amsmath}
\usepackage{multirow}

\newcommand{\newc}{\newcommand}
\newc{\ra}{\rightarrow}
\newc{\lra}{\leftrightarrow}
\newc{\be}{\begin{equation}}
\newc{\ee}{\end{equation}}
\newc{\ba}{\begin{eqnarray}}
\newc{\ea}{\end{eqnarray}}
\newc{\ov}{\overline}
\newc{\pa}{\partial}
\newc{\D}{\Delta}

\def\b{\beta}
\newc{\nn}{\nonumber}

\begin{document}
\thispagestyle{empty}

\hfill CERN-PH-TH/2012-135
\vskip 2truecm
\vspace*{3cm}
\begin{center}
{ {\bf Building $SO(10)$ models from F-theory}}\\
\vspace*{1cm}
{\bf
I. Antoniadis$^{1,\,\flat }$,   G.K. Leontaris$^{2}$}\\
\vspace{4mm}
$^1$  Department of Physics, CERN Theory Division, \\
CH-1211, Geneva 23, Switzerland\vspace{1mm}\\
$^2$ Physics Department, Theory Division, Ioannina University, \\
GR-45110 Ioannina, Greece\vspace{1mm}\\
\end{center}

\vspace*{1cm}
\begin{center}
{\bf Abstract}
\end{center}
\noindent

We revisit local F-theory  $SO(10)$ and $SU(5)$ GUTs and  analyze  their properties within
the framework of the maximal underlying $E_8$ symmetry in the elliptic fibration.
We consider the symmetry enhancements along the intersections of seven-branes with the GUT
surface and study in detail the embedding of the abelian factors undergoing monodromies
in the covering gauge groups. We combine flux data from the successive breaking of
$SO(10)$ to  $SU(5)$ gauge symmetry and subsequently to the Standard Model one, and
further constrain the parameters determining the models' particle spectra.  In order
to eliminate dangerous baryon number violating  operators we propose  ways to construct
 matter parity like symmetries from intrinsic  geometric origin.  We study implementations
 of the resulting constrained scenario in specific examples obtained for a variety of monodromies.

\vfill
$^{\flat}$ {\it On leave from CPHT (UMR CNRS 7644) Ecole Polytechnique, F-91128 Palaiseau, France.}

\newpage

\tableofcontents

\newpage

\section{Introduction}

F-theory~\cite{Vafa:1996xn} provides an interesting `geometrical' reformulation of type IIB superstring theory. It is envisaged that this new approach  will open new possibilities and prove particularly useful to string model building. Indeed,
during the last few  years there is ample evidence that  old successful GUTs including the minimal SU(5),  the SO(10) model etc, are naturally realized in F-theory compactifications~\cite{Donagi:2008ca}-\cite{Blumenhagen:2009up}. The rather interesting fact in F-theory constructions is that because they are defined on a compact elliptically fibered Calabi-Yau complex four dimensional Manifold the exceptional groups $E_6,E_7, E_8$, can be naturally incorporated into the theory too~\cite{Donagi:2008ca,Beasley:2008dc,Beasley:2008kw,Heckman:2009mn}.  Although exceptional gauge symmetries when realized in the context of four-dimensional grand unified theories suffer from several drawbacks, in the case of F-theory models they appear to be more promising as new possibilities arise for the symmetry breaking mechanisms and the derivation of the desired massless spectrum. Indeed, an alternative possibility  is to turn on appropriate $U(1)$ fluxes on the worldvolume of the 7-brane. Applying this mechanism to $SU(5)$ for example it is possible to obtain zero modes
for the Higgs doublets without having dangerous light triplets since they belong to different Dolbeault cohomology groups\cite{Donagi:2008ca,Beasley:2008dc,Beasley:2008kw}. Of course the ultimate goal to build a low energy effective theory with the required spectrum and acceptable quark and lepton mass matrices from  first principles is a formidable task.  Precious tools towards this goal are provided by the twisted eight-dimensional Yang-Mills theory and the particular geometry of the `internal space'   which  determines the wavefunctions'  profiles of the various elementary particles.  This information as well as the successful implementation of the novel GUT breaking mechanism with fluxes on the doublet-triplet problem are convincing  enough to motivate us to delve into the F-theory vacua searching for the ultimate model.

A particularly useful ingredient in building realistic F-theory models, is the notion of
monodromy~\cite{Hayashi:2009ge,Heckman:2009mn,Marsano:2009gv}.
In F-theory configurations involving intersecting seven branes, matter fields appear to localize on the
intersections of two such branes, while the Yukawa couplings occur when three of these branes intersect at a
common point.  Computations of the matter fields wavefunctions~\cite{Hayashi:2009ge,Heckman:2008qa,Font:2008id},
show that these fields are localized at a very tiny region of this intersection.   Monodromy ensures the appearance of tree-level mass for the top quark and  rank-one tree-level fermion mass matrices~\cite{Heckman:2008qa}-\cite{Callaghan:2011jj}, while it plays  a decisive role on
 the final shape of effective models~\cite{Callaghan:2011jj}-\cite{Marsano:2010ix}.  The strength of the corresponding Yukawa couplings are given by integrals of the overlapping wavefunctions picked at the point of the triple
intersection~\cite{Hayashi:2009ge,Heckman:2008qa,Font:2008id,Leontaris:2010zd,Cecotti:2009zf,Aparicio:2011jx,Camara:2011nj,Palti:2012aa}.
\footnote{For similar Yukawa couplings calculations in different context see also~\cite{Cremades:2003qj,Antoniadis:2009bg,Camara:2009xy,Marchesano:2010bs,Krippendorf:2010hj}.}

In the present work we use spectral cover techniques and attempt to thoroughly analyze the  $SO(10)$ as well as  the
(after the symmetry breaking) emerging $SU(5)$  GUT,  with respect to a variety of choices for the monodromy.
In our study we assume that there exists a single point of $E_8$ enhancement in the internal geometry~\cite{Heckman:2009mn} with matter  descending from the adjoint of $E_8$. From the  breaking patterns
$ E_8  \supset SO(10) \times SU(4)_{\perp} $ and $ E_8  \supset  SU(5) \times SU(5)_{\perp}$
it turns out that the spectral covers possess   $SU(4)$ and $SU(5)$ symmetries, respectively.  Thus,
extending  previous works, we classify $SO(10)$ models according to the non-trivial monodromy groups
which are subgroups of the Weyl group $W(SU(4)_{\perp})=S_4$.  Considering in particular the
spectral cover  symmetry breaking to the Cartan subalgebra $SU(4)_{\perp}\to U(1)^3$, we study the quotient
theory after imposing  identifications of these abelian factors under
the subgroups ${\cal Z}_2$, ${\cal Z}_3$, and the Klein four-group ${\cal Z}_2\times {\cal Z}_2$.
 Further, we assume successive breaking
of the   $SO(10)$ symmetry to  $SO(10)\ra SU(5)\times U(1)_X$ through a $U(1)_X$-flux and the
$SU(5)$ to the Standard Model (SM) by a $U(1)_Y$ (hypercharge) flux.  A novel feature of our analysis is the fact
that we impose combined constraints from both $U(1)_{X,Y}$ fluxes to further restrict the flux parameters
of the model.

Using factorization techniques of the spectral cover equations we derive the properties and
in particular the homology classes of the matter curves which constitute an essential tool
towards the computation of flux restrictions and the chiral matter spectrum of the models.
Because monodromies play a particular role on various aspects -including computations
of the Yukawa couplings- we pay particular attention in the way $U(1)$ symmetries are
involved in monodromies. Indeed, matter curves are characterized by the $U(1)$ symmetries
and as a  result, the monodromy action would identify some of them.
 At the next stage of the present work, we combine the obtained data from our analysis and
investigate their implications on the unspecified parameters and  superpotential couplings
 of  viable  effective low energy models.  Among other things, we discuss doublet triplet
 splitting, proton decay and fermion masses. To eliminate dimension-four baryon number violating
 operators we appeal to the geometric properties of the internal manifold and suggest
 ways to construct $Z_2$ discrete symmetries associating them to a possible  matter parity.
 We present examples where such  constructions lead to viable effective low energy models
 and compare them with the existing ones in the recent literature.

More precisely in section 2 we analyze in detail the various monodromies of $SO(10)$
and $SU(5)$ models and the general geometric and flux
constraints imposed on the matter spectrum. In section 3 we elaborate several effective
models and in section 4 we attempt to construct matter parity consistent with
the intrinsic geometry.  We summarize the results and
 present our conclusions in section 5.
 To make this paper self-contained, in the Appendix we  review the necessary F-theory
 tools used in our work,  including  the Weierstrass equation and Tate's forms.
Moreover we  investigate  the gauge symmetry enhancements
 along intersecting matter curves,  and present details of our calculations.

\section{F-theory model building}

F-theory is defined on a background $R^{3,1}\times {\cal X}$  with $R^{3,1}$ the
space-time and ${\cal X}$ an elliptically fibered Calabi-Yau complex fourfold
over a complex three-fold base. In this work we will describe elliptic fibration
through the well known Weierstass model given by the equation~(\ref{Wei}) of
the appendix.
According to the `standard' interpretation  in F-theory the  gauge symmetry is associated
to the singularities of the internal compact manifold. In this work we assume a geometric singularity
 of a two-complex dimension K\"ahler surface  $S_{GUT}$ identified with a group
  $G_S=SO(10)$~\footnote{ For previous studies of
F-$SO(10)$ models see for example~\cite{Beasley:2008dc,Chen:2010ts,Chen:2009me,Donagi:2011dv}.}.
The group structure of the spectral cover is the commutant of $G_S$ to ${\cal E}_8$
\ba
 {\cal E}_8 &\supset & SO(10)\times SU(4)_{\perp}\to SO(10)\times U(1)^3\label{E82U4_1}
 \ea
  where we assumed  further breaking of $SU(4)\to U(1)^3$ by a non-zero adjoint Higgs  vev.
All matter is then found in the representations obtained from the
decomposition of the ${\cal E}_8$-adjoint under the breaking pattern (\ref{E82U4_1})
\ba
248&\ra&(45,1)+(16,4_{\perp})+(\ov{16},\bar 4_{\perp})+(10,6_{\perp})+(1,15_{\perp})
\label{E8adj}
\ea
 Each  $SO(10)$ representation $R$ resides on a matter  curve $\Sigma_{R}$ which
 is distinguished by the specific
charge it carries under the  $U(1)^3$ Cartan subalgebra,
characterized by the $SU(4)_{\perp}$ weights $t_i$
\be
\Sigma_{16}:\; t_i;\; \Sigma_{10}:\; t_i+t_j;\;\Sigma_1: \;\pm (t_i-t_j),\;i\ne j\label{specurves}
\ee
while in addition there are three singlets with zero weight.

Investigating the properties of semi-local models, we can extract useful information for
 the matter curves using Tate's Algorithm~\cite{Tate75}.
In particular for the  $SO(10)$ singularity  the Weierstrass equation -associated to the
elliptic fibration of the internal manifold- takes the form
\be
\label{Wei4}
\begin{split}
y^2&=x^3+b_5xyz+b_4x^2z+b_3yz^2+b_2xz^3+b_0z^5
\end{split}
\ee
with $x,y$ being homogeneous coordinates of the torus fiber and $b_k$ are functions of the
coordinates of the three-fold base. These are sections of $[b_k]=\eta-k t$, where
$\eta=6c_1-t$ with $c_1,-t$ being the first Chern classes of the tangent and normal bundle to
$S_{GUT}$ respectively. The singularity enhancements where chiral matter is
found are studied in detail in the Appendix. Important properties of the local model are
also  `encoded' in the spectral cover equation which for the $SU(4)$ case reads
\ba
{\cal C}_4=\sum_{k=1}^4b_ks^{4-k}&=&0
\label{4sc0}
\ea
with $s$ being an affine parameter.  Furthermore, a detailed analysis summarized
in the Appendix shows that the `locations' of the seven branes associated to
the $SO(10)$ non-trivial representations are determined from the equations
\[ {\bf 16}:\; b_4=0,\;\;\; {\bf 10}:\; b_3^2=0\]

The three additional $U(1)$'s in the breaking pattern of equation (\ref{E82U4_1}) are expected
to put constraints on the superpotential couplings of the effective low energy
model. In particular the simultaneous existence of all three $U(1)$ symmetries would not
allow a  tree-level Yukawa coupling for the top quark -ubiquitous   in all viable  fermion  textures.
This indicates the existence of possible  monodromies  among the
$U(1)$'s  allowing the emergence of a rank one fermion mass matrix structure.
In the present case the possible monodromies are associated to ${\cal Z}_2$, ${\cal Z}_3$
and ${\cal Z}_2\times {\cal Z}_2$  discrete symmetries leading to
the equivalent ${\cal C}_{2+1+1}, {\cal C}_{3+1}, {\cal C}_{2+2}$ factorizations of
the spectral cover. In other words, these imply the identification of two weights $\{t_1,t_2\}$
or three of them $\{t_1,t_2,t_3\}$ or pairing them in two distinct sets $\{t_1,t_2\}, \{t_3,t_4\}$
correspondingly.   Each of these cases leads to
 a different class of low energy models which we examine in the subsequent sections.

\subsection{${\cal Z}_2$ monodromy}

In this case the  factorization of the spectral cover equation is
\ba
{\cal C}_4&=&(a_1+a_2s+a_3s^2)(a_4+a_5s)(a_6+a_7s)\label{z2case_1}
\ea
Comparing this to (\ref{4sc0}) we extract equations of the form
 $b_k =b_k(a_i)$ and use them to derive the relations for
 the homologies $[a_i]$ of the coefficients $a_i$. These are of the form
\ba
\eta-k\,c_1&=&[a_l]+[a_m]+[a_{n}],\;\; {\rm with} \, k+l+m+n=15
\ea
where $l,m,n$ take the values $1,2,\dots, 7$ and $k=0,1,2,3,4$.
Note that we also need to solve the constraint $b_1(a_i)=0$ adopting a suitable Ansatz
for some $a_i$ coefficients. The solution can be expressed in terms of two arbitrary
parameters $\chi_5=[a_5], \chi_7=[a_7]$. We present these details the Appendix.

The above can be used to determine the topological properties of
matter curves. In particular, from the equation $b_4=a_1a_4a_6=0$
we deduce that the three  ${\bf 16}$'s left after the monodromy action are
determined by  $a_1=0$, $a_4=0$ and $a_6=0$. Similarly, the equation
$b_3^2(a_i)=0$ determines the properties of ${\bf 10}$'s.
We collect all the results in Table~\ref{T_ALL} where we introduce for convenience $\chi=\chi_5+\chi_7$.
\begin{table}
\begin{center}
\begin{tabular}{|l|l|l|l|l|}
Matter&$t_i$ charges& Section& Homology &$U(1)_X$\\
\hline
${\bf 16}$&$t_{1,2}$&$a_1$&$\eta-2c_1-\chi$&$M-P$
\\
${\bf 16}$&$t_3$&$a_4$&$-c_1+\chi_5$&$P_5$
\\
${\bf 16}$&$t_4$&$a_6$&$-c_1+\chi_7$&$P_7$
\\
${\bf 10}$&$t_{1,2}+t_3$&$(a_1-\lambda a_4a_6)$&$\eta-2c_1-\chi$&$M-P$
\\
${\bf 10}$&$t_{1,2}+t_4$&$(a_1-\lambda a_4a_6)$&$\eta-2c_1-\chi$&$M-P$
\\
${\bf 10}$&$t_{1}+t_2$&$(a_5a_6+a_4a_7)$&$-c_1+\chi$&$P$
\\
${\bf 10}$&$t_{3}+t_4$&$(a_5a_6+a_4a_7)$&$-c_1+\chi$&$P$
\\
\hline
\end{tabular}
\end{center}
\caption{ Properties of $SO(10)$ representations in the ${\cal Z}_2$ monodromy.}
\label{T_ALL}
\end{table}

\subsubsection{Flux restrictions and multiplicities}

To proceed further,  we investigate the implications of the gauge symmetry
breaking of $ SO(10)$ to $ SU(5)\times U(1)_X$ by a $U(1)_X$ flux.
We introduce the following notation for the $U(1)_X$ flux parameters
\ba
M&=&{\cal F}_1\cdot (\eta-3c_1)\nn\\
P&=&{\cal F}_1\cdot (\chi-c_1)\nn\\
P_n&=&{\cal F}_1\cdot (\chi_n-c_1),\;n=5,7\nn\\
C&=&-{\cal F}_1\cdot c_1\nn
\ea
 This way we obtain the results of the  last column of Table~\ref{T_ALL}.
We should mention that if we wish to protect the $U(1)_X$ boson from
receiving a Green-Schwartz (GS) mass we need to impose
\[ {\cal F}_1\cdot \eta =0,\; \&\; {\cal F}_1\cdot c_1=0\]
which automatically imply $M=C=0$.
In this case, the sum $P=P_5+P_7$ stands for the total flux permeating matter curves
while one can observe form Table~\ref{T_ALL} that the flux
vanishes independently  on the $\Sigma_{16}$ and $\Sigma_{ 10}$ matter curves.

Assuming that $M_{10}^a$  is the number of ${\bf 10}_{t_1+t_3}\in SO(10)$,
 after the $SO(10)$ breaking we obtain the multiplicities
      \ba
{\bf {10}}_{t_1+t_3,-t_2-t_4}&\ra&\left\{\begin{array}{lll}
{5}_{-t_2-t_4}&M_{10}^a+(M-P)&=M_{5_2}\\
{\bar 5}_{t_1+t_3}&M_{10}^a-(M-P)&=-M_{5_1}
  \end{array}\right.
         \ea
Similarly, if  $M_{10}^b$ is the number of ${\bf 10}_{t_3+t_4}\in SO(10)$, we get
     \ba
 {\bf {10}}_{t_3+t_4,-t_1-t_2}&\ra&\left\{\begin{array}{lll}
{5}_{-t_1-t_2}&M_{10}^b+P&=M_{5_{h_u}}\\
{\bar 5}_{t_3+t_4}&M_{10}^b-P&=-M_{5_4}
  \end{array}\right.
         \ea
Now, let $M_{16}^1$ the multiplicity of $16_{t_1}$.
Again from Table~\ref{T_ALL}  we have~\footnote{ Let us clarify that we interpret the
multiplicities obtained from {\bf 16}-decompositions as follows:
 \[\# {\bar 5}_{t_{1,2}+t_5}- \#  { 5}_{-t_{1,2}-t_5}= M_{16}^1-(M-P)\]
and so on.  This is subsequently identified with $-M_{5_3}$, so for the 5's of SU5
we will count: $ M_{5_i}=\#  { 5}-\# {\bar 5} $.
For later use we denote $M_{5_{1,2,3...}}$ the multiplicities of $5's\in SU(5)$.
  Numbering of $10_j,5_i$'s has been chosen to comply with the
  notation $M_{5_i}\dots$ of refs~\cite{Dudas:2010zb,Leontaris:2010zd}.   }
\ba
{\bf 16}_{t_{1,2}}&=&\left\{\begin{array}{lll}
{ 10}_{t_{1,2}}& M_{16}^1&=M_{10_1}\\
         {\bar 5}_{t_{1,2}+t_5}&M_{16}^1-(M-P)&=-M_{5_3}
      \\
         { 1}_{t_{1,2}-t_5}& M_{16}^1+(M-P)&=M_{S_1}
          \end{array}\right.
         \ea
Continuing as above:
\ba
{\bf 16}_{t_{3}}&=&\left\{\begin{array}{lll}
{ 10}_{t_{3}}& M_{16}^3&=M_{10_2}\\
         {\bar 5}_{t_{3}+t_5}&M_{16}^3-P_5&=-M_{5_{h_u}}
      \\
         { 1}_{t_{3}-t_5}& M_{16}^3+P_5&=M_{S_{35}}
          \end{array}\right.
          \\
          {\bf 16}_{t_{4}}&=&\left\{\begin{array}{lll}
{ 10}_{t_{4}}& M_{16}^4&=M_{10_3}\\
         {\bar 5}_{t_{4}+t_5}&M_{16}^4-P_7&=-M_{5_{6}}
      \\
         { 1}_{t_{4}-t_5}& M_{16}^4+P_7&=M_{S_{45}}
          \end{array}\right.
         \ea
where $M_{10_i}, M_{5_j}$ stand for the numbers of $10\in SU(5)$ and $5\in SU(5)$
representations (a negative value corresponds to the conjugate representation).
$M_{S_{ij}}$ denote the multiplicities of the singlet fields. In fact,
as for any other representation, this means that
\[M_{ij}= \# 1_{t_i-t_j}-\# 1_{t_j-t_i}\]
thus, if $M_{ij}>0$ then there is an excess of  $M_{ij}$ singlets $1_{t_i-t_j}=\theta_{ij}$
and vice versa.

\subsubsection{ $SU(5)$ spectrum}

Next we consider the breaking of the   $SU(5)$ symmetry by the hypercharge ($U(1)_Y$) flux.
 The case of  ${\cal Z}_2$  monodromy corresponds to the
following splitting of the spectral cover equation
\ba
b_0\prod_{i}(s-t_i)=(a_1+a_2s+a_3s^2)(a_4+a_5s)(a_6+a_7s)(a_8+a_9s)=\sum_{k=0}^5 b_k s^{5-k}\label{5split_1}
\ea
Following~\cite{Dudas:2010zb} we derive the relations of $b_k(a_i)$ by
equating coefficients of the same powers in $s$, while we use a suitable Ansatz for $a_i$'s to solve
the constraint $b_1(a_i)=0$.
The tenplets are found by studying the zeroth order of the above polynomial, which is
$ b_5=a_1a_4a_6a_8$.
These are designated as
 \ba
 10^{(1)}_{t_1},10^{(2)}_{t_3},10^{(3)}_{t_4},10^{(4)}_{t_5}
 \label{10sZ2}
 \ea
while their homologies are identified as those of $a_1, a_4, a_6, a_8$ respectively.
To determine the properties of the fiveplets we need the corresponding spectral cover equation.
This is a 10-degree polynomial
\[ {\cal P}_{10}(s)\propto\sum_{n=1}^{10}c_ns^{10-n}=b_0\prod_{i,j}(s-t_i-t_j),\; i<j,\; i,j=1,\dots, 5\]
Using~(\ref{5split_1}) we can convert the coefficients $c_n=c_n(t_j)$  to functions of
$c_n(b_j)$.  In particular we are interested for the value ${\cal P}_{10}(0)$
given by the coefficient $c_{10}$ which can be expressed in terms of $b_k$ according to
\ba
c_{10}(b_k)&=& b_3^2b_4-b_2b_3b_5+b_0b_5^2=0\label{c10}
\ea
Using the equations $b_k(a_i)$ and the Ansatz, we can split this equation into seven factors
which correspond to the seven distinct fiveplets left after the ${\cal Z}_2$ monodromy action:
\be
\label{5spitZ2}
\begin{split}
P_5&=  \left(a_4 a_7 a_9+a_5 \left(a_7 a_8+a_6
   a_9\right)\right)\times
   \left(a_1-c a_4 \left(a_7 a_8+a_6 a_9\right)\right)\times \left(a_1-c a_6 \left(a_5 a_8+a_4 a_9\right)\right)\\
   & \times \left(a_1-c \left(a_5 a_6+a_4 a_7\right) a_8\right)\times \left(a_5 a_6+a_4 a_7\right)\times \left(a_5 a_8+a_4
   a_9\right) \left(a_7 a_8+a_6 a_9\right)
   \end{split}
   \ee
These  are assigned as $5^{(0)},5^{(1)},5^{(2)},5^{(3)},5^{(4)},5^{(5)},5^{(6)}$ respectively, while
their homologies can be specified using those of $a_i$ given in Table~\ref{T_0} of the Appendix.
Notice that three of  the above factors correspond to  three fiveplets of the same homology class $[a_1]=\eta-2c_1-\chi$.
The complete spectrum is presented in Table~\ref{SU5Z2spec}.
\begin{table}[tbp] \centering%
\begin{tabular}{|l|c|c|c|c|}
\hline
$SO(10)\supset SU(5)$&$U(1)_i$& homology ($SU(5)$)& $U(1)_Y$-flux&$U(1)_X$\\
\hline
$ 16_{t_1}\supset 10^{(1)}$& $t_{1,2}$& $\eta-2c_1-{\chi}$&$ -N$ &$M_{16}^1$\\ \hline
$ 16_{t_3}\supset 10^{(2)}$& $t_{3}$& $-c_1+\chi_5$&$ N_5$ &$M_{16}^3$\\ \hline
$ 16_{t_4}\supset 10^{(3)}$& $t_{4}$& $-c_1+\chi_7$&$ N_7$ &$M_{16}^4$\\ \hline
$ 45_{t_5}\supset 10^{(4)}$& $t_{5}$& $-c_1+\chi_9$&$ N_9$ &$C$\\ \hline
$10_{t_3+t_4}\supset 5^{(0)}$& $-t_{1}-t_2$& $-c_1+{\chi}$&$ N$ &$M_{10}^b+P$\\ \hline
$ 10_{t_1+t_4}\supset 5^{(1)}$& $-t_{1,2}-t_3$& $\eta -2c_1-{\chi}$&$ -N$ &$-M_{10}^a+M-P$\\ \hline
$10_{t_1+t_4}\supset 5^{(2)}$& $-t_{1,2}-t_4$& $\eta -2c_1-{\chi}$&$ -N$ &$M_{10}^a+M-P$\\ \hline
$ 16_{t_1}\supset 5^{(3)}$& $-t_{1,2}-t_5$& $\eta -2c_1-{\chi}$&$ -N$ &$-M_{16}^1+M-P$\\ \hline
$ 10_{t_3+t_4}\supset 5^{(4)}$& $-t_{3}-t_4$& $-c_1+{\chi}-\chi_9$&$N-N_9$&$-M_{10}^b+P$\\ \hline
$ 16_{t_3}\supset5^{(5)}$& $-t_{3}-t_5$& $-c_1+{\chi}-\chi_7$&$ N-N_7$ &$-M_{16}^3+P_5$\\ \hline
$ 16_{t_4}\supset 5^{(6)}$& $-t_{4}-t_5$& $-c_1+{\chi}-\chi_5$&$ N-N_5$ &$-M_{16}^4+P_7$
\\ \hline
\end{tabular}%
\caption{Field representation content under $SU(5)$, (${\cal Z}_2$ case) their homology class and flux
restrictions under $U(1)_{Y}$. Note that ${\chi}=\chi_5 +  \chi_7 + \chi_9$  so
the $U(1)_Y$ fluxes satisfy $N=N_7 +  N_8 + N_9$. The last column shows the multiplicities
  imposed under $U(1)_{X}$. The $U(1)_X$ fluxes  determine the unspecified $M_i$'s of
   $SU(5)$ model, while  if we assume no chirality in the $SO(10)$ bulk, then $C=0$ and $P_5+P_7=P$.}
\label{SU5Z2spec}
\end{table}

\subsection{${\cal Z}_2\times {\cal Z}_2$ monodromy}

 A second class of models arises from the case  of ${\cal C}_{2+2}$ factorization.
In this case we write  the spectral cover equation as follows
\ba
{\cal C}_4&=& \left(a_3 s^2+a_2 s+a_1\right) \left(a_6 s^2+a_5 s+a_4\right)\label{z2z2case_1}
\ea
This splitting implies the $t_1\lra t_2$ and $t_3\lra t_4$ identifications.
Comparing with the coefficients $b_k$  as in the ${\cal Z}_2$ case we determine the homologies
of $a_i$ given in Table \ref{T_22} of the Appendix and the properties of the matter
curves given in Table \ref{Z22_ALL}.
\begin{table}
\begin{center}
\begin{tabular}{|l|l|l|l|l|}
Matter& Section& Homology &$t_i$ charges&$U(1)_X$\\
\hline
${\bf 16}_a$&$a_1$&$\eta-2c_1-\chi$&$t_{1,2}$&$M-P$
\\
${\bf 16}_b$&$a_4$&$\chi$&$t_{3,4}$&$P-C$
\\
${\bf 10}_a$&$a_1-\lambda a_4$&$\eta-2c_1-\chi$&$t_{1,2}+t_{3,4}$&$M-P$
\\
${\bf 10}_b$&$a_5$&$\chi-c_1$&$t_{3}+t_4/t_1+t_2$&$P$
\\
\hline
\end{tabular}
\end{center}
\caption{ The properties of the $SO(10)$ representations in the ${\cal Z}_2\times {\cal Z}_2$ case.}
\label{Z22_ALL}
\end{table}

Decomposing the $SO(10)$ representations to $SU(5)$ multiplets we get the following
multiplicities.  For the ${\bf 16}$'s
\ba
{\bf 16}_{t_{1,2}}&=&\left\{\begin{array}{lll}
{ 10}_{t_{1,2}}& M_{16}^1&\\
         {\bar 5}_{t_{1,2}+t_5}&M_{16}^1-(M-P)&
      \\
         { 1}_{t_{1}-t_5}& M_{16}^1+(M-P)&
          \end{array}\right.
          \\
          {\bf 16}_{t_{3,4}}&=&\left\{\begin{array}{lll}
{ 10}_{t_{3,4}}& M_{16}^3&\\
         {\bar 5}_{t_{3,4}+t_5}&M_{16}^3-(P-C)&
      \\
         { 1}_{t_{3}-t_5}& M_{16}^3+(P-C)&
          \end{array}\right.
         \ea
The ${\bf 10}$'s decompose to $5_2+\bar 5_{-2}$. The following $SU(5)$ originate
from their decomposition
      \ba
{\bf {10}}_{t_1+t_2,t_3+t_4}&\ra&\left\{\begin{array}{lll}
{5}_{t_1+t_2}&M_{10}^b+P&\\
{\bar 5}_{t_3+t_4}&M_{10}^b-P&
  \end{array}\right.
         \ea
             \ba
{\bf {10}}_{t_1+t_3,t_2+t_4}&\ra&\left\{\begin{array}{lll}
{5}_{t_1+t_3}&M_{10}^a+(M-P)&\\
{\bar 5}_{t_2+t_4}&M_{10}^a-(M-P)&
  \end{array}\right.
         \ea
Because of the $t_1\lra t_2$ and $t_3\lra t_4$ monodromies  the indices  of
the fiveplets resulting from ${\bf 10}_b$ can be identified $t_1+t_3=t_2+t_4$.
Therefore we can assign the same indices while
\[ \# 5_{t_1+t_3}-\# \bar 5_{t_1+t_3} = 2 (M-P)\]

\subsubsection{$SU(5)$ spectrum}

In the ${\cal Z}_2\times {\cal Z}_2$ case, the spectral cover equation for $SU(5)$ obtains the form
\[{\cal C}_{5}(s)=  \left(a_3 s^2+a_2 s+a_1\right) \left(a_6 s^2+a_5 s+a_4\right) (a_7+a_8 s)\]
Proceeding as in the $SO(10)$ case, we identify the relations
$b_k(a_i), k=1,\dots 5$ by comparing coefficients of the same power in $s$.
The homology classes of $a_i$ are given in Table~\ref{22HC}.

The $10\in SU(5)$ are obtained from the solutions of the equation   $b_5= a_1a_4a_7=0$
therefore they are associated to $a_1=0, a_4=0$ and $a_7=0$.
The fiveplets are found  by solving  the corresponding equation (\ref{c10})
and  in terms of the $a_i$'s we can write
\be
\label{P5SU5}
\begin{split}
P_5&= (a_6 a_7 + a_5 a_8) \times (a_1^2 - a_1 (a_5 a_7 + 2 a_4 a_8) \lambda +
   a_4 (a_6 a_7^2 + a_8 (a_5 a_7 + a_4 a_8)) \lambda^2)\\
&\times (a_1 -
   a_5 a_7 \lambda) \times (a_6 a_7^2 + a_8 (a_5 a_7 + a_4 a_8))
  \times a_5  \end{split}
   \ee
   This equation has  five factors corresponding to an equal number of fiveplets
  dubbed as \[5^{(0)},5^{(1)},5^{(2)},5^{(3)},5^{(4)}\]
    in the order of appearance  in the above product.  It is straightforward to determine
    their  homology classes  using the results of table \ref{22HC} in the Appendix. To compute
    the flux     restrictions we define
  \[{\cal F}_Y\cdot {\psi}=-N_1-N_2,\;
  {\cal F}_Y\cdot {\chi}=N_2,\; {\cal F}_Y\cdot \eta ={\cal F}_Y\cdot c_1=0\]
  The results are summarized in Table \ref{SU5Z2Z2}.
   \begin{table}[tbp] \centering%
\begin{tabular}{|l|c|c|c|c|}
\hline
$SO(10)\supset SU(5)$&$U(1)_i$& homology ($SU(5)$)& $U(1)_Y$-flux&$U(1)_X$\\
\hline
$ 16_{t_1}\supset 10^{(1)}$& $t_{1,2}$& $\eta-2c_1-{\chi}-\psi$&$ N_1$ &$M_{16}^1$\\ \hline
$ 16_{t_3}\supset 10^{(2)}$& $t_{3}$& $-2c_1+\chi$&$ N_2$ &$M_{16}^3$\\ \hline
$10_{t_3+t_4}\supset 5^{(0)}$& $-t_{1}-t_2$& $-c_1+{\chi}+\psi$&$ -N_1$ &$M_{10}^b+P$\\ \hline
$10_{t_1+t_4}\supset 5^{(1)}$& $-t_{1,2}-t_3$& $2\eta -4c_1-2{\chi}-2\psi$&$ 2N_1$ &$2(M-P)$\\ \hline
$ 16_{t_1}\supset 5^{(2)}$& $-t_{1,2}-t_5$& $\eta -2c_1-{\chi}-\psi$&$ N_1$ &$-M_{16}^1+M-P$\\ \hline
$ 16_{t_3}\supset5^{(3)}$& $-t_{3,4}-t_5$& $ -2c_1+{\chi}+\psi$&$ -2N_1-N_2$ &$-M_{16}^3+P-C$\\ \hline
$ 10_{t_3+t_4}\supset 5^{(4)}$& $-t_{3}-t_4$& $-c_1+\chi$&$N_2$&$-M_{10}^b+P$\\ \hline
$ 45\supset 10^{{(3)}}$& $t_{5}$& $-c_1+\psi$&$-N_1-N_2$&$M_{10}^c$\\ \hline
\end{tabular}%
\caption{Field representation content under $SU(5)$, their homology class and flux
restrictions under $U(1)_{Y}$ for the ${\cal Z}_2\times {\cal Z}_2$ case.}
\label{SU5Z2Z2}
\end{table}

\subsection{${\cal Z}_3$ monodromy}

We write the spectral cover equation as follows
\ba
{\cal C}_4&=&\left(a_1+a_2 s+a_3 s^2+a_4 s^3\right) \left(a_5+s a_6\right)\label{z3case_1}
\ea
Comparing with the coefficients $b_k$, we determine as previously the
homologies of $a_i$'s.  We may choose two different ways to
solve the $b_1=0$ constraint, either   $a_5=\lambda a_6,\, a_3=-\lambda a_4$
 or $a_3=-\lambda a_5,\, a_4=\lambda a_6$.
The homology classes of $a_i$ are shown in Table \ref{T_Z3} of the Appendix.

Proceeding as in the previous cases we determine the homologies of ${\bf 16},\,{\bf 10}$ matter
curves. In particular from the equation $b_4=a_1a_5$ we infer that the homologies of
${\bf 16}$'s are  $\eta-3c_1-\chi$ and $\chi-c_1$ respectively, while  for those
of the ${\bf 10}$'s we choose them in consistency of the defining equation
$b_3^2=0$. The results are collected in Table~\ref{T16Z3}.
\begin{table}[t]
\begin{center}
\begin{tabular}{|l|l|l|l|l|}
Matter& Section& Homology &$t_i$ charges&$U(1)_X$\\
\hline
${\bf 16}$&$a_1$&$\eta-3c_1-\chi$&$t_{i}$&$M-P+C$
\\
${\bf 16}$&$a_5$&$\chi-c_1$&$t_{4}$&$P$
\\
${\bf 10}$&$a_1+\lambda a_2$&$\eta-3c_1-\chi$&$2t_{i}$&$M-P+C$
\\
${\bf 10}$&$a_6$&$\chi$&$t_{i}+t_4$&$P-C$
\\
\hline
\end{tabular}
\caption{${\cal Z}_3$ case $A$. Properties of $SO(10)$ representations  (with $i$ any of $1,2,3$).}
\label{T16Z3}
\end{center}
\end{table}
For the second  Ansatz the properties of the ${\bf 10}$-representations change according to
Table~\ref{T16Z3B}.
\begin{table}[t]
\begin{center}
\begin{tabular}{|l|l|l|l|l|}
Matter& Section& Homology &$t_i$ charges&$U(1)_X$\\
\hline
${\bf 16}$&$a_1$&$\eta-3c_1-\chi$&$t_{i}$&$M-P+C$
\\
${\bf 16}$&$a_5$&$\chi-c_1$&$t_{4}$&$P$
\\
${\bf 10}$&$a_1a_6+a_2a_5$&$\eta-3c_1$&$2t_{i}$&$M$
\\
${\bf 10}$&$a_1a_6+a_2a_5$&$\eta-3c_1$&$t_{i}+t_4$&$M$
\\
\hline
\end{tabular}
\caption{${\cal Z}_3$ case $B$. Properties of $SO(10)$ representations  (with $i$ any of $1,2,3$).}
\label{T16Z3B}
\end{center}
\end{table}

\subsubsection{$SU(5)$ spectrum}

In this case the relevant spectral cover polynomial  is
\[\sum_{k=0}^5 b_ks^{5-k}=\left(a_4 s^3+a_3 s^2+a_2 s+a_1\right) \left(a_5+s a_6\right) \left(a_7+s a_8\right)\]
We can easily extract the equations  determining the coefficients
$b_k(a_i)$, while the corresponding one  for the homologies reads
\[[b_k]=\eta-k c_1= [a_l]+[a_m]+[a_n],\;k=0,1,\dots,5,\; k+l+m+n=18, \;l,m,n\le 8\]
 The solution $b_1=0$  and other details are shown in the Appendix.
The tenplets are determined by $b_5=a_1 a_5 a_7=0$, while
the equation  (\ref{c10}) factorizes as follows
\be
\label{Z35split}
\begin{split}
P_5&=\left(a_1 a_6 a_8+a_2 \left(a_6 a_7+a_5 a_8\right)\right)\times
\left(a_1 a_6+a_5 \left(a_2-c a_5 a_7\right)\right)\\
&\times \left(a_7 \left(a_2-c a_5 a_7\right)+a_1 a_8\right)\times
   \left(a_6 a_7+a_5 a_8\right)
\end{split}
\ee
The four factors determine the homologies of the fiveplets dubbed $5^{(0)},5^{(1)},5^{(2)},5^{(3)}$
 correspondingly. These, together with the tenplets are given in Table~\ref{SU5Z3spec}.
   \begin{table}[t] \centering%
\begin{tabular}{|l|c|c|c|c|}
\hline
$SO(10)\supset SU(5)$&$U(1)_i$& homology ($SU(5)$)& $U(1)_Y$-flux&$U(1)_X$\\
\hline
$ 16_{t_i}\supset 10^{(1)}$& $t_{i}$& $\eta-3c_1-{\chi}-\psi$&$- N_x-N_y$ &$M_{16}^1$\\ \hline
$ 16_{t_4}\supset 10^{(2)}$& $t_{4}$& $-c_1+\chi$&$ N_x$ &$M_{16}^2$\\ \hline
$10_{t_i+t_4}\supset 5^{(0)}$& $-2t_{i}$& $\eta-3c_1$&$ 0$ &$M_{10}+n\,M$\\ \hline
$10_{t_i+t_j}\supset 5^{(1)}$& $-t_{i}-t_4$& $\eta -3c_1-{\psi}$&$ -N_y$ &$-M_{10}+n\,M$\\ \hline
$ 16_{t_i}\supset 5^{(2)}$& $-t_{i}-t_5$& $\eta -3c_1-{\chi}$&$- N_x$ &$-M_{16}^1+(M-P+C)$\\ \hline
$ 16_{t_4}\supset5^{(3)}$& $-t_{4}-t_5$& $-c_1+{\chi}+\psi$&$ N_x+N_y$ &$-M_{16}^2+P$\\ \hline
$ 45\supset 10^{{(3)}}$& $t_{5}$& $-c_1+\psi$&$N_y$&$M_{10}'$\\ \hline
\end{tabular}%
\caption{Field representation content under $SU(5)$, their homology class and flux
restrictions with respect to $U(1)_{Y}$ for the ${\cal Z}_3$ case. The coefficient $n$ takes
the values 1 for case $A$ and 2 for case $B$.}
\label{SU5Z3spec}
\end{table}

Decomposition of ${\bf 10}$s of $SO(10)$ in more subtle.  Observe first that $10_{t_k+t_4}$ and
 eventually their complex conjugates $\ov{10}_{-t_k-t_4}$ sit on a matter curve (dubbed here $\Sigma_{10_1}$).
 Also $10_{t_i+t_j}$ and their complex conjugates are on  $\Sigma_{10_2}$. Notice however that
 $t_i+t_j+t_k+t_4=0$, so that  $\ov{10}_{-t_k-t_4}=10_{t_i+t_j}$, i.e., the two curves accommodate
fields with the same $U(1)_{t_a}$ charges.

Adopting $t_{1,2,3}=t_i$ we have
\ba
{\bf {10}}_{t_1+t_3}\ra {\bf {10}}_{2t_i,\,-t_i-t_4}&\ra&\left\{\begin{array}{lll}
{5}_{-t_i-t_4}&M_{10}^a+Q&\\
{\bar 5}_{2t_i}&M_{10}^a-Q&
  \end{array}\right.
\ea
where $Q=(M-P+C)$ for the first model and $Q=M$ for the second one.
Also
             \ba
{\bf {10}}_{t_1+t_4}\ra {\bf {10}}_{t_1+t_4,\,-2t_i}&\ra&\left\{\begin{array}{lll}
{5}_{-2t_i}&M_{10}^b+U&\\
{\bar 5}_{t_i+t_4}&M_{10}^b-U&
  \end{array}\right.
  \ea
  with $U=P-C$ and $U=M$ respectively.

         At the $SU(5)$ level we have only two distinct species of fiveplets
         emerging from ${\bf 10}$ of $SO(10)$ and the differences are given
         ($M_{10}=M_{10}^b-M_{10}^a$)
\ba
\#  {5}_{-2t_i}-\# {\bar 5}_{2t_i}&=& M_{10}+V \nn\\
\#  {5}_{-t_i-t_4}-\# {\bar 5}_{t_i+t_4}&=& -M_{10}+V \nn
         \ea
where $V=Q+U=n\, M$  with $n=1,2$ for the first and second case respectively.

\section{Low energy effective models}

Once we have obtained the properties of matter curves and the necessary ingredients
we can start our investigation for a viable effective low energy theory emerging from the
above picture. To discuss low energy effective models of $SO(10)$ origin, we further need to
specify the symmetry breaking mechanism down to the Standard Model.  We will
assume here that $SO(10)$  breaks to $SU(5)$ by flux along $U(1)_X$.
At the next stage of symmetry breaking a $U(1)_Y$ flux must
 be turned on.   In the subsequent we need to take into account the hypercharge splitting.
 When $SO(10)$ breaks by fluxes, we have already seen  how the $U(1)_X$ flux restricts on
 the relevant matter curves providing information for the chiral $SU(5)$ spectrum.
 In a similar way, $U(1)_Y$ flux splits the $SU(5)$ multiplets.

Assuming that the integer $N_{Y}$ represents the effect of the  $U(1)_{Y}$ hypercharge
flux on a specific matter curve  and that the $U(1)_{X}$ flux is given by the integers
$M_{10,5}$ the resulting SM spectrum appears with the following multiplicities:
\ba
{\bf 10}\in SU(5)\Rightarrow\left\{\begin{array}{lll}
n_{(3,2)_{\frac 16}}-n_{(\bar 3,2)_{-\frac 16}}&=&M_{10}\\
n_{(\bar 3,1)_{-\frac 23}}-n_{(3,1)_{\frac 23}}&=&M_{10}-N_Y\\
n_{(1,1)_{1}}-n_{(1,1)_{-1}}&=&M_{10}+N_Y
\end{array}
\right.
\label{10dec}
\ea
and
\ba
{\bf 5}\in SU(5)\Rightarrow\left\{\begin{array}{lll}
n_{(3,1)_{-\frac 13}}-n_{(\bar 3,1)_{\frac 13}}&=&M_5\\
n_{(1,2)_{\frac 12}}-n_{(1, 2)_{-\frac 12}}&=&M_{5}+N_Y
\end{array}
\right.
\label{5dec}
\ea
In our subsequent analysis we will use these formulae repeatedly for each matter curve to determine the SM matter spectrum.

\subsection{${\cal Z}_2$ models}

For the case of ${\cal Z}_2$  monodromy we have already given the properties of the $SO(10)$
spectrum in Table~\ref{T_ALL}.
   All properties of the $SU(5)$ representations are summarized in Table~\ref{SU5Z2spec}.

  In previous studies in the context of the $SU(5)$ level, the $M_{5_i},M_{10_j}$ are
   not fully  determined.  It is a fact however  that  there is a non-trivial connection between the
  $U(1)_i$ symmetries and the flux.  One condition on $M_i$'s can be
 imposed by demanding   the $U(1)$'s  to be traceless.
 This implies~\cite{Dudas:2010zb,Marsano:2010sq}
  \ba
  \sum M_{10_j}+\sum_i M_{5_i}=0\label{DPrel}
  \ea
Interestingly, this condition can be recovered by appealing to the properties of a specific  $U(1)$.
  In particular if we assume the embedding of $SU(5)$ in $SO(10)\supset SU(5)\times U(1)_X$,
  the associated $U(1)_X$ flux implies  restrictions on $M_i$'s. The relation (\ref{DPrel}) can be
  derived  under the following two assumptions:
   \be
   {\cal F}_1\cdot \eta=0\; {\rm and }\; {\cal F}_1\cdot c_1=0\label{MC0}
   \ee
which imply that both flux parameters $M,C$ are zero  $M=C=0$. Note that these conditions
are equivalent to those imposed to the $U(1)_Y$ flux. As in that case,
 they ensure that the $U(1)_X$ boson does not get a GS term. Thus,
adding all $M_i's$ from Table~\ref{SU5Z2spec} while respecting the above conditions  we get
\be
\begin{split}
 \sum_iM_i&= 0\nn
\end{split}
\ee
since $P=P_5+P_7$ and $M=0$. Thus, this is the equivalent to the relation  (\ref{DPrel}).
Notice however that $U(1)_X$ flux imposes additional relations among $M_i$'s as can be
seen from  the last column of Table~\ref{T_ALL}.

Our main goal in this section is to present specific models with realistic properties and spectra,
combining the flux constraints emerging from the successive flux breaking.
In order  to assess the implications of these combined data,
we will compare our findings with the $SU(5)$ models already discussed in the literature.

 A class of models may arise as follows: We have seen that assuming a ${\cal Z}_2$ monodromy,
 the matter curves accommodating ${\bf 16}$'s reduce to three and there are four
 curves left over to accommodate the ${\bf 10}$'s of $SO(10)$.
 Further breaking of the gauge group down
 to $SU(5)$ with fluxes, will result to three matter curves
  $\Sigma_{10_{t_1}},\Sigma_{10_{t_3}},\Sigma_{10_{t_4}}$ accommodating the $10\in SU(5)$
  and seven  curves  $\Sigma_{5_{t_i+t_j}}$ for the relevant $5$'s.
   A fourth curve could also accommodate a $10\in SU(5)$ from the bulk adjoint ${\bf 45}$, however
  we assume that no chirality is generated in the $SO(10)$ bulk. These are shown in Table \ref{SU5Z2spec}.

We consider the case where  at least the third family resides on $\Sigma_{10_{t_1}}$ and demand
that the mass is obtained at tree level. Then the up-Higgs is found in   $\Sigma_{5_{t_1+t_2}}$
\[\lambda_t\, 10_{t_1}\cdot 10_{t_2}\cdot 5_{-t_1-t_2}\;\stackrel{{\cal Z}_2}\longrightarrow \; \lambda_t\, 10_{t_1}\cdot 10_{t_1}\cdot 5_{-2t_1}
\]
Since up and bottom quarks constitute a left handed doublet they naturally reside on the same matter curve.
Demanding that the bottom quark has also a tree-level coupling, the only available is
\[\lambda_b\, 10_{t_1}\cdot \bar 5_{t_2+t_5}\cdot \bar 5_{t_3+t_4}\;\stackrel{{\cal Z}_2}\longrightarrow \;
\lambda_b\, 10_{t_1}\cdot \bar 5_{t_1+t_5}\cdot \bar 5_{t_3+t_4}\]
where one of the fiveplets $\bar 5_{t_1+t_5}, \bar 5_{t_3+t_4}$ should accommodate
the right handed  chiral matter and the other the down-type Higgs field.

Now, having in mind the properties of the  $\Sigma_{10_{t_i}}$ curves with respect the
$U(1)_Y$-flux (see relevant column of table \ref{SU5Z2spec}) we find that a convenient choice of
multiplicities is
 \[M_{16}^1=4,M_{16}^3=0,M_{16}^4=-1\]
 These ensure that there exist three $Q=(u,d)^T$
doublets to accommodate the corresponding fermionic degrees of freedom and an
additional vector like pair $Q+\ov{Q}$. Furthermore the choice
\[N=N_9=0,\; N_5=1,\; N_7=-1\]
will result to the following matter content:
\ba
{\bf 10}_{t_{1}}&=&\left\{\begin{array}{lll}
n_{(3,2)_{1/6}}-n_{(\bar 3,2)_{-1/6}}=4&\ra&4\times Q\\
 n_{(\bar 3,1)_{-2/3}}-n_{( 3,1)_{2/3}}=4&\ra&4\times u^c
      \\
n_{(1,1)_{1}}-n_{(1,1)_{-1}}=4&\ra& 4\times e^c
          \end{array}\right.
         \ea

\ba
{\bf 10}_{t_{3}}&=&\left\{\begin{array}{lll}
n_{(3,2)_{1/6}}-n_{(\bar 3,2)_{-1/6}}=0&\ra&0\times Q\\
 n_{(\bar 3,1)_{-2/3}}-n_{( 3,1)_{2/3}}=-1&\ra&1\times\bar u^c
      \\
n_{(1,1)_{1}}-n_{(1,1)_{-1}}=1&\ra& 1\times e^c
          \end{array}\right.
         \ea

\ba
{\bf 10}_{t_{4}}&=&\left\{\begin{array}{lll}
n_{(3,2)_{1/6}}-n_{(\bar 3,2)_{-1/6}}=-1&\ra&1\times\overline{Q}\\
 n_{(\bar 3,1)_{-2/3}}-n_{( 3,1)_{2/3}}=0&\ra&0\times\bar u^c
      \\
n_{(1,1)_{1}}-n_{(1,1)_{-1}}=-2&\ra& 2\times \bar e^c
          \end{array}\right.
         \ea
 The first matter curve   $ 10^{(1)}$  accommodates four complete $10_{t_1}$'s of $SU(5)$.
 On the second   curve  $ 10^{(2)}$ we obtain $\overline{Q}$ from an incomplete $\ov{10}_{-t_3}$
 and only an $e^c$ state from the corresponding $10_{t_3}$. Finally,  $ 10^{(3)}$ provides
 $\ov{Q}+2\bar e^c$ from $\ov{10}_{t_4}$'s.
Therefore, we end up with $3\times (Q+u^c+e^c)$, one pair $Q+\overline{Q}$
and $u^c+\bar u^c$ and two pairs $e^c+\bar e^c$. The spectrum is
shown in Table~\ref{restZ2}.

The choice of the flux parameters should secure the existence of three
$5_f$'s and the Higgs doublets $h_u+h_d$. A constraint emerges then from
the fact that  the number of triplets minus antitriplets should be $-3$.
To this end we impose
\[(M_{10}^b+P)+(M-P-M_{10}^a)+(M_{10}^a+M-P)+(M-P-4)+(P-M_{10}^b)+(P_5
+1+P_7)=-3\]
which is automatically satisfied for $M=C=0$.  Notably, this is equivalent to the assumptions
in (\ref{MC0}).
\begin{table}[tbp] \centering%
\begin{tabular}{|l|c|c|c|c|}
\hline
 $SU(5)$ curves &$U(1)_i$&  $U(1)_Y$-flux&$U(1)_X$\\
\hline
$\Sigma_{10^{(1)}}$& $t_{1,2}$&$ 0$ &$4$\\ \hline
$ \Sigma_{10^{(2)}}$& $t_{3}$& $ 1$ &$0$\\ \hline
$ \Sigma_{10^{(3)}}$& $t_{4}$& $ -1$ &$-1$\\ \hline
$\Sigma_{10^{(4)}}$& $t_{5}$&$ 0$ &$0$\\ \hline
$ \Sigma_{5^{(0)}}$& $-t_{1}-t_2$& $ 0$ &$M_{10}^b+P$\\ \hline
$ \Sigma_{5^{(1)}}$& $-t_{1,2}-t_3$& $ 0$ &$-M_{10}^a+M-P$\\ \hline
$ \Sigma_{5^{(2)}}$& $-t_{1,2}-t_4$&$ 0$ &$M_{10}^a+M-P$\\ \hline
$\Sigma_{5^{(3)}}$& $-t_{1,2}-t_5$& $ 0$ &$-4+M-P$\\ \hline
$ \Sigma_{5^{(4)}}$& $-t_{3}-t_4$&$0$&$-M_{10}^b+P$\\ \hline
$\Sigma_{5^{(5)}}$& $-t_{3}-t_5$& $ 1$ &$P_5$\\ \hline
$ \Sigma_{5^{(6)}}$& $-t_{4}-t_5$& $ -1$ &$1+P_7$\\ \hline
\end{tabular}%
\caption{Field representation with three families and extra vector like pairs
$Q+\bar Q$, $u^c+\bar u^c$ and two $e^c+\bar e^c$ in ${\cal Z}_2$ case.}
\label{restZ2}
\end{table}
Next  we use the freedom to adjust the remaining flux parameters $M_{10}^i,P_j$
to obtain the appropriate $5$ and $\bar 5$ representations. There are several
options and in the subsequent analysis we discuss a few of them.

\subsubsection{Model  $A$:}

We start our investigation by imposing a condition to ensure the  up-type Higgs $h_u$.
Since we need at least one $h_u$, the simplest possibility is
\[M_{10}^b+P=1\;\ra\;M_{10}^b=1-P \]
while we take $M=C=N=M^a_{10}=0$.
Because $N=0$, the Higgs fiveplet living on $5^{(0)}$ does not split.
The same happens also for fiveplets  eventually residing on $ 5^{(1,2,3,4)}$ curves.
Splitting occurs only on matter curves $ 5^{(5,6)}$. In particular, we have

\ba
{ 5}^{(5)}=\left\{\begin{array}{l}
n_{(3,1)_{-1/3}}-n_{(\bar 3,1)_{1/3}}=P_5\\
 n_{(1,2)_{1/2}}-n_{(1,2)_{-1/2}}=P_5+1
          \end{array}\right.,\;
          { 5}^{(6)}=\left\{\begin{array}{l}
n_{(3,1)_{-1/3}}-n_{(\bar 3,1)_{1/3}}=1+P-P_5\\
 n_{(1,2)_{1/2}}-n_{(1,2)_{-1/2}}=P-P_5
          \end{array}\right.
         \ea
Another constraint arises from the fact that the down-Higgs  should
reside in $\bar 5^{(4)}$
\[2 P-1<0\ra P\le 0\]
Assuming the minimal case $P=0$, we get
\[\bar 5^{(4)}\ra 1\times (\bar D, h_d)\]
Table \ref{rest1} summarizes the spectrum where the only parameter left unspecified
up to now is $P_5$.  Different choices of $P_5$ affect the way $\Sigma_{ 5^{(5,6)}}$
split leading to different effective models.

\begin{table}[tbp] \centering%
\begin{tabular}{|l|c|c|c|c|}
\hline
curves &$U(1)_i$& SM Matter content\\
\hline
$\Sigma_{ 10^{(1)}}$& $t_{1,2}$&$4\times (Q+u^c+e^c)$\\ \hline
$ \Sigma_{10^{(2)}}$& $t_{3}$&$1\times (\bar u^c+e^c)$\\ \hline
$\Sigma_{\overline{10}^{(3)}}$& $t_{4}$& $1\times \overline Q+2\times \bar e^c$\\ \hline
$\Sigma_{10^{(4)}}$& $t_{5}$&$0$\\ \hline
$ \Sigma_{5^{(0)}}$& $-t_{1}-t_2$&$1\times (h_u+D)$\\ \hline
$\Sigma_{ 5^{(1)}}$& $-t_{1,2}-t_3$& $0$\\ \hline
$\Sigma_{ 5^{(2)}}$& $-t_{1,2}-t_4$&$0$\\ \hline
$\Sigma_{5^{(3)}}$& $-t_{1,2}-t_5$&$4\times (d^c+\ell)$\\ \hline
$ \Sigma_{5^{(4)}}$& $-t_{3}-t_4$&$1\times (\bar D+h_d)$\\ \hline
$\Sigma_{5^{(5)}}$& $-t_{3}-t_5$& $P_5\times D+(1+P_5)\times h$\\ \hline
$\Sigma_{ 5^{(6)}}$& $-t_{4}-t_5$& $(1-P_5)\times D+(-P_5)\times h$\\ \hline
\end{tabular}%
\caption{Case A of ${\cal Z}_2$: Field representation of  $SU(5)$  matter curves (first column).
Last column shows the multiplicities of the SM representations.}
\label{rest1}
\end{table}

The following Yukawa couplings emerge involving the $10$'s:
\ba
 10^{(1)} 10^{(1)} 5^{(0)}&\ra& Q\,u^c\,h_u
 \\
  10^{(1)} {10}^{(2)} 5^{(0)}\theta_{13}&\ra&u^c\,e^c\,D \langle\theta_{13}\rangle
  \\
  10^{(1)} \ov{10}^{(2)} \theta_{31}&\ra&\bar u^cu^c\,
  \langle\theta_{31}\rangle  \\
  10^{(1)} \ov{10}^{(3)} \theta_{41}&\ra&(\ov{Q}Q+\bar e^ce^c)
  \langle\theta_{41}\rangle
  \ea
Choosing $P_5=1$, the last two 5's give
\ba
5^{(5)} &=&1\times D+2\times h_u'\\
\bar 5^{(6)} &=&0\times \bar D+1\times h_d'
\ea
In this case, splitting occurs on 5-curves other than the Higgs.
The triplets can in principle form the following mass terms
\[{\cal W}_T\propto (\langle\theta_{15}\rangle{ \bar 5^{(3)}}+\langle\theta_{13}\rangle\langle\theta_{14}\rangle{\bar 5^{(4)}})\,{ 5^{(0)}}+
(\langle\theta_{31}\rangle{\bar 5^{(3)}}+\langle\theta_{54}\rangle{\bar 5^{(4)}})\, { 5^{(5)}}\]
and the doublets
\[{\cal W}_{h_{u,d}}\propto (\langle\theta_{15}\rangle{\bar 5^{(3)}}+\langle\theta_{13}\rangle\langle\theta_{14}\rangle{\bar 5^{(4)}}+\langle\theta_{15}\rangle\langle\theta_{14}\rangle{\bar 5^{(6)}})\, { 5^{(0)}}+
(\langle\theta_{31}\rangle{\bar 5^{(3)}}+\langle\theta_{54}\rangle\bar { 5^{(4)}}+\langle\theta_{34}\rangle{\bar 5^{(6)}})\, { 5^{(5)}}\]
The Higgs doublets mass matrix differs from that of the color triplets, allowing for the possibility
to make the latter massive while keeping one Higgs doublet pair light. However, a
 detailed knowledge of Yukawa couplings and possible fine tuning
would be needed to make triplets massive while keeping doublets light.
Therefore, we proceed to a second set of flux parameters.

\subsubsection{Model $B$:}

We can choose the down Higgs doublet to descend from a $\Sigma_5$-curve whose matter
context  splits by flux and derive the bottom coupling from a suitable non-renormalizable (NR)-term.

 Choosing in this case $M_{10}^a=M_{10}^b=P=1, P_7=-1,P_5=2$, the spectrum is that of Table
\ref{rest2}.
 \begin{table}[tbp] \centering%
\begin{tabular}{|l|c|c|c|c|}
\hline
curves &$U(1)_i$& SM Matter content\\
\hline
$ \Sigma_{10^{(1)}}$& $t_{1,2}$&$4\times (Q+u^c+e^c)$\\ \hline
$ \Sigma_{10^{(2)}}$& $t_{3}$&$1\times (\bar u^c+e^c)$\\ \hline
$\Sigma_{\overline{10}^{(3)}}$& $t_{4}$& $1\times \overline Q+2\times \bar e^c$\\ \hline
$\Sigma_{10^{(4)}}$& $t_{5}$&$0$\\ \hline
$\Sigma_{ 5^{(0)}}$& $-t_{1}-t_2$&$2\times (D,h_u)$\\ \hline
$\Sigma_{ 5^{(1)}}$& $-t_{1,2}-t_3$& $2\times (\bar D,h_d')$\\ \hline
$\Sigma_{ 5^{(2)}}$& $-t_{1,2}-t_4$&$0$\\ \hline
$\Sigma_{5^{(3)}}$& $-t_{1,2}-t_5$&$5\times (d^c+\ell)$\\ \hline
$\Sigma_{ 5^{(4)}}$& $-t_{3}-t_4$&$0$\\ \hline
$\Sigma_{5^{(5)}}$& $-t_{3}-t_5$& $2\times D+3\times h_u'$\\ \hline
$\Sigma_{ 5^{(6)}}$& $-t_{4}-t_5$& $0\times D+1\times h_d$\\ \hline
\end{tabular}%
\caption{Case B of ${\cal Z}_2$: Field representation of  $SU(5)$  matter curves (first column).
Last column shows the multiplicities of the SM representations.}
\label{rest2}
\end{table}
The bottom receives a mass
\[10^{(1)}\bar 5^{(3)}\bar  5^{(6)}\theta_{35}\]
The triplets can in principle form the following mass terms
\[{\cal W}_T\propto (\langle\theta_{15}\rangle{ \bar 5^{(3)}}+\langle\theta_{13}\rangle{\bar 5^{(1)}})\,{ 5^{(0)}}+
(\langle\theta_{31}\rangle{\bar 5^{(3)}}+\langle\theta_{51}\rangle{\bar 5^{(1)}})\, { 5^{(5)}}\]
while  the doublets involve additional terms,
\[{\cal W}_{h_{u,d}}\propto (\langle\theta_{15}\rangle{\bar 5^{(3)}}+\langle\theta_{13}\rangle\rangle{\bar 5^{(1)}}+\langle\theta_{15}\rangle\langle\theta_{14}\rangle{\bar 5^{(6)}})\, { 5^{(0)}}+
(\langle\theta_{31}\rangle{\bar 5^{(3)}}+\langle\theta_{51}\rangle\bar { 5^{(1)}}+\langle\theta_{34}\rangle{\bar 5^{(6)}})\, { 5^{(5)}}\]
A suitable choice of $\theta_{ij}$ vevs can leave one Higgs doublet pair light, while all triplets receive
heavy masses.

\subsubsection{Model  $C$  :}

We present now a model with the three families distributed in three different matter curves.
We choose $10^{(1)},10^{(2)},10^{(3)}$ to accommodate the $(Q,u^c,e^c)$ SM multiplets, thus
$M_{16}^i=1,\,i=1,3,4$. Further we take $N=N_9=0$ and $N_7=-N_5=1$. From these arrangements,
we get a complete tenplet $10^{(1)}=(Q,u^c,e^c)$ whose content we identify with the third generation.
Because of the non-trivial flux, the remaining two tenplets decompose to pieces with
different multiplicities. The $10^{(2)}$ in particular gives
\be
10^{(2)}\Rightarrow\left\{\begin{array}{ll}
n_{(3,2)_{1/6} }  - n_{(\overline 3 ,2)_{ - 1/6} } &  = 1\times Q\\
n_{(\overline 3 ,1)_{ - 2/3} }  - n_{(3,1)_{2/3} } & = 2\times u^c\\
n_{(1,1)_1 }  - n_{(1,1)_{ - 1} }  & = 0
\end{array}
\right.
\ee
and the $10^{(3)}$ representation content is split according to
\be
10^{(2)}\Rightarrow\left\{\begin{array}{ll}
n_{(3,2)_{1/6} }  - n_{(\overline 3 ,2)_{ - 1/6} } &  = 1\times Q
\\
n_{(\overline 3 ,1)_{ - 2/3} }  - n_{(3,1)_{2/3} } & = 0
\\
n_{(1,1)_1 }  - n_{(1,1)_{ - 1} }  &  = 2\times e^c
\end{array}
\right.
\ee
\begin{table}[tbp] \centering%
\begin{tabular}{|l|c|c|c|c|c|c|c|}
\hline
 Matter curve &$SU(5)\times U(1)_i$&  $N_Y$&$M_X$&SM reps&$R$-parity\\
\hline
$\Sigma_{ 10^{(1)}}$& $10_{t_{1,2}}$&$ 0$ &$1$&$1\times (Q,t^c,\tau^c)$&$-$\\ \hline
$\Sigma_{ 10^{(2)}}$& $10_{t_{3}}$& $ -1$ &$1$&$1\times Q+2\times u^c$&$-$\\ \hline
$\Sigma_{ 10^{(3)}}$& $10_{t_{4}}$& $ 1$ &$1$&$1\times Q+2\times e^c$&$-$\\ \hline
$\Sigma_{ 5^{(0)}}$& $5_{-t_{1}-t_2}$& $ 0$ &$3$&$3\times (D,h_u)$&$+$\\ \hline
$\Sigma_{ 5^{(1)}}$& $\bar 5_{t_{1,2}+t_3}$& $ 0$ &$-1$&$1\times (d^{c'},\ell')$&$-$\\ \hline
$\Sigma_{ 5^{(2)}}$& $\bar 5_{t_{1,2}+t_4}$&$ 0$ &$-1$&$1\times (b^{c},\ell_{\tau})$&$-$\\ \hline
$\Sigma_{5^{(3)}}$& $\bar 5_{t_{1,2}+t_5}$& $ 0$ &$-2$&$2\times (\overline{ D},h_d')$&$-$\\ \hline
$\Sigma_{ 5^{(4)}}$& $\bar 5_{t_{3}+t_4}$&$0$&$-1$&$1\times (d^c,\ell)$&$-$\\ \hline
$\Sigma_{5^{(5)}}$& $\bar 5_{t_{3}+t_5}$& $ -1$ &$0$&$1\times h_d$&$+$\\ \hline
$\Sigma_{ 5^{(6)}}$& $\bar 5_{t_{4}+t_5}$& $ 1$ &$-1$&$1\times \overline{ D}$&$-$\\ \hline
\end{tabular}%
\caption{Field representation for ${\cal Z}_2$-case $C$ of ${\cal Z}_2$ monodromy.}
\label{restC}
\end{table}
We choose $5^{(0)}=5_{h_u}$ to accommodate the up Higgs so that the coupling
\[{\cal W}\supset 10^{(1)}10^{(1)}5^{(0)} \; \Rightarrow\;\lambda_t\,10_{3}10_{3}5_{h_u}\]
provides mass to the top-quark.

In the present case we take advantage of the possibility to realize the doublet-triplet
splitting and simultaneously  obtain a tree-level mass for the
bottom quark, by assigning the down type Higgs to $\bar 5^{(5)}=\bar 5_{h_d}$ and
 making the following choice
\[ P=P_5=1,\,P_7=0,\, M^a=0,\, M^b=2\]
The resulting spectrum is shown in Table~\ref{restC}.

Then  the relevant available coupling
 $10^{(1)}\bar 5^{(2)}\bar 5^{(5)}$ implies that the third generation
 can be assigned to $\bar 5^{(2)}=\bar 5_3$, thus
\[ {\cal W}\supset  10^{(1)}\bar 5^{(2)}\bar 5^{(5)}\; \Rightarrow\;\lambda_b\,10_{3}\bar 5_3\bar 5_{h_d}\]
However, in addition to the above the symmetries of the model allow also  the couplings
$10^{(1)}\bar 5^{(1)}\bar 5^{(6)}$ and  $10^{(1)}\bar 5^{(3)}\bar 5^{(4)}$, involving
dangerous 4-d proton decay  operators.

 A standard way to avoid such terms   is to impose
matter parity. In F-theory compactifications this is possible by engineering a $Z_2$-symmetry
on both the Calabi-Yau fourfold ${\cal X}$ and the $G^{(4)}$ flux~\cite{Hayashi:2009bt}.
This symmetry is communicated to the matter curves so that the
massless spectrum splits to positive and negative parity eigenstates. Provided a suitable
$Z_2$ action is imposed on ${\cal X}, G^{(4)}$ pair, baryon violating and other
unwanted terms can be eliminated as long as they do not conserve the induced $R$-parity.
Choosing the $R$-parity of the various massless fields as in the last column of Table
\ref{restC}, we can easily check that all the dangerous terms are eliminated.

We remark that this model has exactly the same chiral matter spectrum with the
model presented in~\cite{Leontaris:2010zd}.  The requirement that this present version has  an
$SO(10)$ origin  has only led to an augmented  Higgs sector by two more
Higgs fiveplet pairs which can be massive by an appropriate coupling. In particular
the following Yukawa couplings
\be
{\cal W}\supset \langle \theta_{15}\rangle \,\bar 5^{(3)}\,5^{(0)}+
\langle \theta_{14}\rangle\langle \theta_{15}\rangle \bar 5^{(6)}\,5^{(0)}
\ee
give heavy masses to the Higgs triplets.  The model predicts a reasonable
fermion mass hierarchy as discussed in~\cite{Leontaris:2010zd}. Thus, we
will not pursue it further here.

\subsection{${\cal Z}_2\times {\cal Z}_2$ models}

As noted above, compared to the previous (${\cal Z}_2$) case, now  the number of
curves is further reduced. Therefore, the possibilities of distributing matter
on different curves are reduced and models arising in this context are eventually
more constrained. The restrictions notwithstanding do not exclude a variety of models.
We start by picking up the numbers
\be
 M_{16}^1=3, M_{16}^3=0,  M_{10}^c=0, M_{10}^b=1, N_1=0, N_2=-1, M=0, P=0, C=0\label{Z2Z2Inte}
 \ee
Following the same procedure as above, we derive the SM spectrum shown in Table~\ref{Z2Z2ex}.
In addition, there are singlets $\theta_{ij}, i,j=1,2,3,4$ on the spectral curves
defined in (\ref{specurves})  as well as singlets $\theta_{i5}$ from the ${\bf 16}$ decompositions.
 The latter singlets in particular, appear on line bundles on the surface $S_{GUT}$
and $U(1)$ fluxes breaking the gauge group are affecting their chirality and multiplicities.
Thus, we have
\be
\label{single}
\begin{split}
\#\, {\rm of} \,\theta_{15}-\#\, {\rm of} \,\theta_{51} &= M_{16}^1+(M-P)=3\\
\#\, {\rm of} \,\theta_{35}-\#\, {\rm of} \,\theta_{53}&= M_{16}^3+P-C=0
\end{split}
\ee
The results (\ref{single}) tell us that there is an excess of $\theta_{15}$ over $\theta_{51}$ while
possible $\theta_{35}/\theta_{53}$  arise only in pairs. Depending on the
detailed geometrical properties of the compact space,  these  pairs can be
either massive or massless.  In the latter case they are expected
to  receive heavy masses and decouple.

The $SO(10)$ Yukawa couplings are:
\[{\bf 16}_{t_1}\cdot{\bf 16}_{t_1}\cdot{\bf 10}_{t_3+t_4}\ra 10^{(1)}10^{(1)}5^0+
 10^{(1)}\bar 5^{(2)}\bar 5^{(4)}+1_{t_1-t_5}\bar 5^{(2)}\, 5^{(0)}\]
Identifying $ 10^{(1)}=10_M$, $\bar 5^{(2)}=\bar 5_M$, $5^0=5_{h_u}$,  $\bar 5^{(4)}=\bar 5_{h_d}$
and $1_{t_1-t_5}=\theta_{15}$ we observe that all SM fermion masses arise from this coupling
\[10_M 10_M 5_{h_u},\; 10_M \bar 5_M \bar 5_{h_d},\; \theta_{15} \bar 5_M 5_{h_u}\]
To avoid large Matter-Higgs  mixing through the fiveplets, we take $\langle\theta_{15}\rangle =0$.
\begin{table}[tbp] \centering%
\begin{tabular}{|l|c|c|c|c|c|}
\hline
 $SO(10)\, \supset \,SU(5)$ &$U(1)_i$& $N_Y$&$U(1)_X$&SM reps&$R$-parity \\
 \hline
$ 16_{t_1}\supset 10^{(1)}$& $t_{1,2}$&$ 0$ &$3$&$3\times (Q,u^c,e^c)$&$-$\\ \hline
$ 16_{t_3}\supset 10^{(2)}$& $t_{3}$& $ -1$ &$0$&$1\times (u^c, \bar e^c)$&$-$\\ \hline
$10_{t_3+t_4}\supset 5^{(0)}$& $-t_{1}-t_2$& $0$ &$1$&$1\times (D, h_u)$&$+$\\ \hline
$10_{t_1+t_4}\supset 5^{(1)}$& $-t_{1,2}-t_3$& $ 0$ &$0$&$0$&$-$\\ \hline
$ 16_{t_1}\supset \bar 5^{(2)}$& $t_{1,2}+t_5$& $ 0$ &$-3$&$3\times (d^c,\ell)$&$-$\\ \hline
$ 16_{t_3}\supset5^{(3)}$& $-t_{3,4}-t_5$& $1$ &$0$&$1\times h_u'$ &$-$\\ \hline
$ 10_{t_3+t_4}\supset \bar 5^{(4)}$& $t_{3}+t_4=2t_3$&$-1$&$-1$&$1\times\bar D, 2\times h_d $&$+$\\ \hline
$ 45\supset 10^{{(3)}}$& $t_{5}$&$1$&$0$&$ 1\times (\bar u^c, e^c)$&$-$\\ \hline
\end{tabular}%
\caption{Field representation content under $SU(5)$,  and  flux choices
 for the ${\cal Z}_2\times {\cal Z}_2$ case.  Neutral singlet fields $\theta_{ij}$ are not
 shown in this table (see text). }
\label{Z2Z2ex}
\end{table}

  When the $U(1)_{X,Y}$ fluxes are turned on, $\Sigma_{16_{t_3}}$ curve contains a $u^c$-type state descending
  from $16_{t_3}$ and a $\bar e^c\in \ov{16}_{t_3}$.
Because  $16_{t_3}\ra 10^{(2)}+$nothing, while due to the absence of $5^{(1)}$ (see table~\ref{Z2Z2ex}) the coupling
\[{\bf 16}_{t_1}{\bf 16}_{t_3}{\bf 10}_{t_1+t_4}\ra  \{ 10^{(1)}+ \bar 5^{(2)}+
1_{t_1-t_5}\}\cdot\{10^{(2)}\} \cdot \{{\rm nothing}\}\]
does not provide any $SU(5)$ invariant coupling.

Furthermore, the coupling
\be
{\bf 45}\cdot{\bf 16}_{t_3}\cdot {\bf\ov{16}}_{-t_3}\ra 10^{(3)}\ov{10}^{(2)}\theta_{53}\label{Exmass}
 \ee
 could in principle  make extra matter massive,  but
the realization of such a coupling would  require a  $\theta_{53}$-vev from an additional
 $\theta_{35}/\theta_{53}$-pair provided that this pair is massless.  The  coupling
\be
\theta_{13}^2 \, 5^{(0)}\bar 5^{(4)}\ra \langle\theta_{13}^2 \rangle \,D\,\ov{D}\label{Trimass}
\ee
makes extra triplets massive, whilst the  coupling
\be
(\theta_{13}^2 \,\bar 5^{(0)}+\theta_{53} \, 5^{(3)}) \bar 5^{(4)}\label{Doumass}
\ee
could make massive one linear combination  of $h_{u}$ and one $h_d$  doublets.

\subsubsection{Minimal variation}

The above minimal version appears to be a promising scenario provided there is a
way to make extraneous matter massive. One of the possible issues here is
that it requires a singlet vev which appears only in pairs $\theta_{53}/\theta_{35}$
and without knowing the detailed geometry we cannot claim whether it is massless or not.
We could replace this by other combinations, like
$\theta_{54}\theta_{43}$ etc. Another way is to modify some of the flux parameters.

Thus, in addition to the  non-zero  integers in (\ref{Z2Z2Inte}), let us assume
also non-zero values for
\[M_{16}^3=-1, M_{10}^c=C=+1\]
Then, the number of $5^{(3)}$'s does not change but the number of singlets becomes
\be
\#\, {\rm of} \,\theta_{35}-\#\, {\rm of} \,\theta_{53}= M_{16}^3+P-C=-2\label{singlets}
\ee
thus, we have two copies of $\theta_{53}$.  There are modifications on the exotic spectrum too.
In particular
\[10^{(2)} \ra  1\times \bar Q'+ 2\bar e^c,\; 10^{(3)} \ra  1\times  Q'+ 2 e^c\]
The modified massless spectrum is given in Table~\ref{Z2Z2ex1}.

We discuss now the modifications induced on Yukawa couplings.
Because $\theta_{53}$ is now present in the spectrum, the coupling (\ref{Doumass})
is now suitable to provide a mass to a pair $h'_uh'_d$.
However, the chiralities of the representations involved in (\ref{Exmass}) are reversed.
 To make extra matter massive we need a coupling of the form
 \be
 \ov{10}^{(2)}_{-t_3}10_{t_5}\Phi_{t_3-t_5}\label{Mexotics}
 \ee
where $\Phi_{t_3-t_5}$ an effective singlet to compensate for a pissible
 absence of a massless $\theta_{35}$. Indeed, we could have for example
\[\Phi_{t_3-t_5}=\theta_{31}\theta_{15}\]
The $SO(10)$ origin of (\ref{Mexotics}) is
\[{\bf\ov{16}}_{-t_3}\cdot{\bf 45}\cdot{\bf 16}_{t_1}\,\theta_{31}\]
Notice that $\theta_{31}$ is  a singlet even at the $SO(10)$ level, since its origin
is from the spectral cover $SU(4)$ adjoint
\[15\ra \sum_{i\ne j}^4{\bf 1}_{t_i-t_j}+ 3 \times {\bf 1}_0 \ra \theta_{ij}+3\times \theta_0\]
\begin{table}[tbp] \centering%
\begin{tabular}{|lcl|c|c|c|c|}
\hline
 $SO(10)$ & $\supset$ & $SU(5)$ &$U(1)_i$& $N_Y$&$U(1)_X$&SM reps \\
 \hline
$ 16_{t_1}$&$\supset$ & $ 10^{(1)}$& $t_{1,2}$&$ 0$ &$3$&$3\times (Q,u^c,e^c)$\\ \hline
$ \ov{16}_{-t_3}$ &$\supset$ &$\ov{10}^{(2)}$& $-t_{3}$& $ -1$ &$-1$&$1\times \bar Q',\, 2\times\bar e^{c'}$\\ \hline
$10_{t_3+t_4}$&$\supset$ &$5^{(0)}$& $-t_{1}-t_2$& $0$ &$1$&$1\times (D, h_u)$\\ \hline
$10_{t_1+t_3}$&$\supset$& $5^{(1)}$& $-t_{1,2}-t_3$& $ 0$ &$0$&$0$\\ \hline
$ 16_{t_1}$&$\supset$& $\bar 5^{(2)}$& $t_{1,2}+t_5$& $ 0$ &$-3$&$3\times (d^c,\ell)$\\ \hline
$ 16_{t_3}$&$\supset$& $5^{(3)}$& $-t_{3,4}-t_5$& $1$ &$0$&$1\times h_u'$ \\ \hline
$ 10_{t_3+t_4}$&$\supset$& $\bar 5^{(4)}$& $t_{3}+t_4=2t_3$&$-1$&$-1$&$1\times\bar D, 2\times h_d $\\ \hline
$ 45$&$\supset$& $10^{{(3)}}$& $t_{5}$&$1$&$1$&$ 1\times Q',\,2\times e^{c'}$\\
 \hline
$ 16_{t_1}$&$\supset$ & $ 1^{(1)}$& $t_{1}-t_5$&$ 0$ &$3$&$3\times \theta_{15}$\\ \hline
$ \ov{16}_{-t_3}$&$\supset$ & $ 1^{(3)}$& $-t_{3}+t_5$&$ -1$ &$-1$&$2\times \theta_{53}$\\
\hline
\end{tabular}%
\caption{A modified field representation content under $SO(10)\supset SU(5)$,
 for the ${\cal Z}_2\times {\cal Z}_2$ case. Singlets arising from $16$'s are also included. (There
 are additional singlets from $SU(4)$ spectral cover not include here, see text)}
\label{Z2Z2ex1}
\end{table}
Before closing this section, we take the opportunity to add a few more comments with regard to the
singlet fields: We have already seen that $U(1)_X$ flux leaves an excess of $\theta_{53}$ fields
over $\theta_{35}$, the specific number of them being determined in terms of the integers
related to the appropriate flux.
In general there may also be vector-like pairs $\theta_{53}+\theta_{35}$, massless or massive depending on
the specific internal geometry.   These singlets appear on line bundles on the surface $S_{GUT}$
and $U(1)$ fluxes breaking the gauge group are affecting their chirality and multiplicities
in the way described above. However, the properties of singlets that do not localize on line bundles
on $S_{GUT}$ are not always known.
 In general, it is observed that chirality is also present in the singlet field spectrum as
 long as these singlet fields
reside on matter curves on $S_{GUT}$.  If  a similar  flux  mechanism is operative at a preceding symmetry level,
it could also eliminate some of $\theta_{ij}, i,j=1,2,3,4$.  As we have already mentioned at the $SO(10)$ level
 $\theta_{13}$ and $\theta_{31}$ are $SO(10)$ singlets and therefore are found on line bundles
 extending normal to  $S_{GUT}$ and away from the local point on $S_{GUT}$.  Although at our
 present level of understanding we do not know the details of the global geometry and the breaking
 mechanism away from $S_{GUT}$ we can possibly deduce such properties~\cite{Callaghan:2011jj} near $S_{GUT}$
 from the spectral cover.
 In the above analysis we assume the existence of both fields since as we  have seen both are needed
 to give mass to extraneous matter fields,
otherwise   one would need a contrived way to get rid of these states.

\subsection{ ${\cal Z}_3$ models}

Finally, in this section  we present two characteristic
examples of effective models when we quotient the theory with
a ${\cal Z}_3$ monodromy.

\subsubsection{ First example}

We make the following choice
\[M_{16}^1= 4, M_{16}^2=-1, M= -1, P= -1, C= 3,M_{10}= 2\,.\]
Then, the required combinations take the values
\[M_{10}+M=+1,\;-M_{10}+M=-3,\;-M_{16}^1+(M-P+C)=-1,\; -M_{16}^2+P=0\,.                   \]
The massless spectrum is presented in Table~\ref{SU5Z3specA_1}. Remarkably, this model is the
$SO(10)$ upgrade of the $3+1+1+1$ example given in~\cite{Dudas:2010zb}. Here, in addition
we are also able to obtain the exact singlet field spectrum whose properties are determined
by the $U(1)_X$ flux. Notice that the singlets $\theta_{ij}$ with $i,j=1,2,3$ because
of the identification $t_1=t_2=t_3$ under ${\cal Z}_3$, carry no charge and therefore
they  do not couple to the fluxes. Further, the multiplicities of $\theta_{i4},\theta_{4i}$
 are not determined by  the $U(1)_{X,Y}$  fluxes assumed here, hence they are treated
 as free parameters.

   \begin{table}[t] \centering%
\begin{tabular}{|l|c|c|l|c|}
\hline
$SO(10)\supset SU(5)$& $N_Y$-flux&$M_X$&Matter&$R$-parity\\
\hline
$ 16_{t_i}\supset 10^{(1)}_{t_i}=10_M$&$+1$ &$4$&$4Q+3u^c+5e^c$&$-$\\ \hline
$ \ov{16}_{-t_4}\supset \ov{10}^{(2)}_{-t_4}=\ov{10}_x$&$ -1$ &$-1$&$1\ov{Q}+2\bar e^c$&$-$\\ \hline
$10_{-2t_i}\supset 5^{(0)}_{-2t_i}=5_{h_u}$&$ 0$ &$+1$&$h_u+D$&$+$\\ \hline
$10_{t_i+t_4}\supset \bar 5^{(1)}_{t_i+t_4}=\bar 5_M$& $ 0$ &$-3$&$3(\ell+d^c)$&$-$\\ \hline
$ 16_{t_i}\supset \bar 5^{(2)}_{t_{i}+t_5}$& $+1$ &$-1$&$\ov{D}$&$-$\\ \hline
$ 16_{t_4}\supset\bar 5^{(3)}_{t_{4}+t_5}$& $ -1$ &$0$&$h_d$&$+$\\ \hline
$ 16_{t_i}\supset 1_{t_i-t_5}$&$+1$ &$4$&$7\times\theta_{15}$&$-$\\ \hline
$ \ov{16}_{-t_4}\supset \ov{1}_{t_5-t_4}$&$ -1$ &$-1$&$2\times \theta_{54}$&$-$\\ \hline
$ 45\supset 10^{{(3)}}_{t_5}$&$0$&$0$&empty&$-$\\ \hline
\end{tabular}%
\caption{Field representation content under $SU(5)$, their homology class and flux
restrictions under $U(1)_{Y}$ for the first example of the ${\cal Z}_3$ case.}
\label{SU5Z3specA_1}
\end{table}
\subsubsection*{$SO(10)$ origin of couplings}
The top Yukawa can be obtained from
\[{\bf 16}_{t_i} {\bf 16}_{t_i}{\bf 10}_{-2t_i} \to 10_M \,10_M\, 5_{h_u} + 10_M \,\bar 5^{(2)}_{t_{i}+t_5}\,\bar 5_M
   + \theta_{15}\,\bar 5^{(2)}_{t_{i}+t_5}\, 5_{h_u}\]
   The first term is indeed the up-quark mass matrix, while the third one provides mass to the extra
   triplet pair through a non-zero vev  $M_D=\langle\theta_{15}\rangle$:
   \[  \langle\theta_{15}\rangle\,\bar 5^{(2)}_{t_{i}+t_5}\, 5_{h_u}\to M_D \bar D\,D\]
Notice that according to our choices, $10_{-2t_i}$ contributes  $M_{10}^b-4$ multiplets of $5_{h_u}$'s, thus we
need to assume $M_{10}^b>4$. Since $M_{10}^b-M_{10}^a=2$, it turns out that $M_{10}^a>2$.  Taking
$M_{10}^a=3$, then $10_{2t_1}$ gives $M_{10}^a-3=0$ of $\bar 5_{2t_1}$'s and the coupling
\[ {\bf  16}_{t_i}{\bf 16}_{t_4}{\bf 10}_{2t_i} \to 10_M \,\bar 5^{(3)}_{t_4+t_5} \not\bar 5_{2t_1}\]
is not realized because $\bar 5_{2t_1}\not\in  10_{2t_1}$.
The $SO(10)$ origin of the bottom coupling is
\ba
{\bf 16}_{t_i}\,{\bf 10}_{t_j+t_4}\,{\bf 16}_{t_4}\,\theta_{k4}&\ra&  10_{M}\,\bar 5_M\, \bar 5_{h_d}\theta_{14},\; \{i,j,k\}=\{1,2,3\}
 \label{mbZ3}
 \ea
A Higgs mixing term would require two $SU(5)$ singlets obtained from
\ba
\theta_{14}\,{\bf 16}_{t_i}\, {\bf 16}_{t_4}\,{\bf 10}_{-2t_j}&\to& \theta_{14}\,\theta_{15}\,\bar 5_{h_d}\,5_{h_u}
\label{HMZ3}
\ea
From the bottom mass term, we infer that $\langle\theta_{14}\rangle$ should be large enough. Similarly
the triplet mass requires also  a non-zero $\langle\theta_{15}\rangle$. These requirements induce
unacceptably large Higgs mixing. However, it is possible to keep triplets light without disturbing
the RGE running~\cite{Blumenhagen:2008aw}-\cite{Davies:2012vu}. In this case we could assume  a small $\langle\theta_{15}\rangle$ vev. Alternatively, we could impose matter parity to eliminate this term.
We discuss this issue in conjunction  with the requirements to avoid possible proton decay operators.
Notice first that $h_u, h_d$ reside on different matter curves, while the triplet $\bar 5_{h_d}$
has been washed away by flux, therefore,  the relevant tree-level graph mediated by the triplet cannot be generated.
Even if massive KK-modes of the above states are considered, they could not form a direct term.
Therefore, proton decay graphs are expected to be suppressed.

Notice also that the operator
\[{\bf 16}_{t_i}\,{\bf 16}_{t_i}\,{\bf 16}_{t_i}\,{\bf 10}_{t_i+t_4} \to 10_{t_i}\,10_{t_i}\,10_{t_i}\,5_{t_i+t_4} \]
would require a non-zero $\theta_{5i}$-vev which does not exist.

Further, we see that the  $SO(10)$ coupling
\[ \theta_{14}{\bf 10}_{-2t_1}{\bf 10}_{t_1+t_4}\to \langle\theta_{14}\rangle\,5_{h_u}\bar 5_M\]
would imply unacceptable mixing among color extra triplets and ordinary matter.

As illustrated in~\cite{Hayashi:2009bt}, provided a suitable
$Z_2$-symmetry is imposed on ${\cal X}, G^{(4)}$ pair, baryon violating and other
unwanted terms can be eliminated if they do not conserve the induced $R$-parity.
Choosing the $R$-parity of the various massless fields as in the last column of Table
\ref{SU5Z3specA_1}, we can easily check that all the dangerous terms are eliminated.
Notice also that the mere existence of the bottom Yukawa coupling (\ref{mbZ3}), requires that
the $SO(10)$ singlet $\theta_{14}$ should be assigned with positive ($+$) $R$-parity. This makes impossible the
existence of the  Higgs mixing term (\ref{HMZ3}), thus  the vev of $\theta_{15}$ can be
chosen at will.

\subsubsection{Second example}

For the second Ansatz of the ${\cal Z}_3$ case, we choose  $N_x=-N_y=1$, $M_{16}^1=3$, $M_{10}=1$, $P=-3$
and $M=C=0$, to obtain a rather interesting  model with the spectrum presented in Table~\ref{SU5Z3specB}.
   \begin{table}[t] \centering%
\begin{tabular}{|l|c|c|l|c|}
\hline
$ SU(5)$& $N_Y$-flux&$M_X$&Matter&$R$-parity\\
\hline
$ 10^{(1)}_{t_i}=10_M$&$0$ &$3$&$3 (Q+u^c+e^c)$&$+$\\ \hline
$5^{(0)}_{-2t_i}=5_{h_u}$&$ 0$ &$+1$&$h_u+D$&$+$\\ \hline
$\bar 5^{(1)}_{t_i+t_4}=\bar 5_{\bar D}$& $ +1$ &$-1$&$\ov{D}$&$\pm$\\ \hline
$\bar 5^{(2)}_{t_{i}+t_5}=\bar 5_{h_d}$& $-1$ &$0$&$h_d$&$\pm$\\ \hline
$ \bar 5^{(3)}_{t_{4}+t_5}=\bar 5_M$& $0$ &$-3$&$3(d^c,\ell)$&$-$\\ \hline
$  [\theta_{14}\cdot \bar 5^1]$&  &&&$+$\\ \hline
$ [\theta_{15}\cdot \bar 5^2]$&  &&&$-$\\ \hline
$ [\theta_{15}\cdot \theta_{14}]$&  &&&$-$\\ \hline
\end{tabular}%
\caption{Field representation content under $SU(5)$, their homology class and flux
restrictions under $U(1)_{Y}$ for the second example of the ${\cal Z}_3$ case.
Last column shows the $R$-parity assignment  used to eliminate unwanted operators.}
\label{SU5Z3specB}
\end{table}
There is also extraneous matter coming in pairs $\bar e^ce^c,\bar u^cu^c$
from the $10^{2,3}$ representations which can be massive
by an appropriate coupling $\sim \langle X\rangle\,(\bar e^ce^c+\bar u^cu^c)$.

The couplings providing with masses the charged fermions and the Higgs triplets are
\[ 10^{1} 10^{1}5^{0}+ 10^{1}\bar 5^{3} \bar 5^{2}\theta_{15}  + 5^0\bar 5^1\theta_{14}
\to 10_M10_M5_{h_u}+10_M\bar 5_M\bar 5_{h_d}+\langle \theta_{14}\rangle\, 5_{h_u}\bar 5_{\bar D}\]
Notice that the bottom mass originates from a fourth order NR-term, thus the corresponding
singlet vev  should be substantially large $\langle\theta_{15}\rangle\gtrsim  10^{-1}M_S$, with
$M_S$ being the GUT-scale.
Consequently, the coupling $5^{0}\bar 5^{2}\theta_{15}$ involving the same vev $\langle \theta_{15}\rangle$
should  be avoided to protect Higgs doublets from receiving a large mass. In addition
dimension four and five proton decay operators $10_M\bar 5_M\bar 5_M\theta_{14}\theta_{15}$ and
$10_M10_M10_M\bar 5_M$  allowed by gauge symmetry, should be eliminated.
To this end we again  appeal to the matter parity which for the present model is chosen according
to the last column of Table~\ref{SU5Z3specB}.  We observe that the above requirements fix only
the product $\theta_{14}\bar 5^1$ to positive,  and $\theta_{15}\bar 5^2, \theta_{14}\theta_{15}$
to negative $R$-parity.   This choice is also compatible with the neutrinos.
Left handed neutrino components living in $\ell=(\nu,e)$
of $\bar 5_M$ should be paired up with right-handed components $\nu^c$, which at the SM gauge
symmetry level should generate mass terms of the form
\[ {\cal W}_{\nu}\sim \lambda_{\nu} h_u\, \ell\,\nu^c+M\,\nu^c\nu^c\]
If we assign matter parity $(+)$ to the $\nu^c$ state,
the first term which gives a Dirac mass could be contained in a non-renormalizable coupling
of the form $\bar 5_M\,5_{h_u}\nu^c\langle\theta_{14}\theta_{15}\rangle$. The role of
the right-handed neutrino could be played by the zero modes $\theta_{ii}$ which are neutral
under the $U(1)$ factors. Alternatively, the right-handed neutrinos can be the Kaluza-Klein
modes as in~\cite{Antoniadis:2002qm} and its F-theory extension~\cite{Bouchard:2009bu}.
We remind that we have left an ambiguity  in the determination
of matter parities of the corresponding states in Table~\ref{SU5Z3specB}.
In the next section, we will see how this is fixed from the intrinsic geometry considerations.

\section{Matter Parity from Geometry}

In several cases of the models above, in order to prevent proton decay
operators  we have appealed to matter parity that might arise from the
internal  geometry and the fluxes. The implementation of this idea
requires a thorough study of the manifold and the flux properties along
the lines of the discussion started in~\cite{Hayashi:2009bt}~\footnote{For other
scenarios to implement such symmetries in F-theory models see~\cite{Ludeling:2011en,Davies:2012vu}.
The importance of deriving such symmetries of string origin has also 
been discussed  recently in \cite{Lee:2011dya} and in the context of Gepner 
model in~\cite{Ibanez:2012wg}. }.

A simpler bottom-up approach to incarnate such symmetries in a local model
could be described as follows. We  consider the GUT divisor $S_{GUT}$  which  locally is
covered by open patches $U_a\in S_{GUT}$.
We focus on a single trivialization patch  and take $s$ to be the coordinate along
the fiber. Here we will relax other possible constraints~\cite{Hayashi:2009bt}
and simply demand  that under the required geometric transformation the
spectral cover equation  should remain invariant up to an overall phase.
To this end consider the transformation $\sigma$ where $s, b_k$ are mapped according to
\[s(\sigma(p))\;=\; s(p)\,e^{\i\phi},\; b_k(\sigma(p))\;=\; b_k(p)\,e^{i(\xi-(6-k)\phi)}\]
Then each term in the spectral cover equation transforms the same way
\[ b_k s^{5-k}\ra e^{i(\xi-\phi)}b_k s^{5-k}\]

This invariance allows two different  ways to communicate a $Z_2$ symmetry to $S_{GUT}$:

For $\phi=0$
\[s\ra s,\; b_k\ra b_k \,e^{i\xi}\]

For a $Z_N$ symmetry we take
\[\phi=\frac{2\pi}{N}\]
thus, for $N=2$, we have $\phi=\pi$  and
\ba
s\ra -s,\; b_k\ra (-1)^{k} e^{i\xi}\,b_k\label{bphase}
\ea

One can think of various ways to implement this idea to a particular
model.  We can construct for example a matter parity based on the
above considerations by  extending  this to the line bundles associated
to the matter and Higgs representations of $SU(5)$.
To do this, we need to know the particular way that these
properties are induced to the corresponding  wavefunctions.
We should remark here that in several cases of the subsequent explorations
the resulting matter parity can differ from the conventional one. Yet,
we will see that these  cases can lead to viable effective models.

\subsection{ A $Z_2$ parity for the ${\cal Z}_2\times {\cal Z}_2$ monodromy}

 For example, a reasonable way to communicate consistently this parity to the
various states residing on the  matter curves is through the coefficients $a_n$
and their relations implied by the splitting of the spectral cover.
 Consider for example the case of $SU(5)$ for ${\cal Z}_2\times {\cal Z}_2$.
 Then the $b_k$'s are given in terms of $a_m$'s by relations of the form
\[ b_k=\sum a_la_ma_n,\; l+m+n=N-k\]
where $N=17$.  If $a_n$  transform as
\[a_n\ra a_n\, e^{i(\zeta  -n\phi)}\]
then
\[ b_k\propto  a_la_ma_n \ra a_la_ma_n \, e^{3\zeta-(N-k)\phi}\]
and for $\phi=\pi$  the phase  of $b_k$ is
\[ (-1)^{k+1}\,e^{i 3\zeta}\]
which is consistent with (\ref{bphase}) for example if $\xi=\pi, \zeta= 0$, thus
\[ a_n(\sigma(p))=a_n(p)\, e^{-in\pi}\]
We still do not know how to correlate these phases to the particular properties
of the line bundles and the wavefunctions associated to them, however there are many
options which we now discuss. We can rely for
example on the fact that the bundles are associated to particular products
of the coefficients $a_n$  being sections of the sums of the latter.
  The $5^{(0)}$ fiveplet for example has a defining equation involving
  the combinations $a_6a_7+a_5a_8$. Under the preceding transformation,
  it gives an overall phase
\[ 5^{(0)}\sim a_6a_7 \ra (-1)^{13}=-1\]
If we associate these phases to the $R$-parity then the representations transform according to:
\[10^{(i)}(-),\; 5^{(0)}(-),\; 5^{(1)}(+),\; 5^{(2)}(-),\; 5^{(3)}(+),\; 5^{(4)}(-)\]
However, this is not the parity used in Table \ref{Z2Z2ex} thus either another
identification of the matter spectrum should be used or a more sophisticated parity
construction should be associated to the matter curves.

\subsection{ The case of  ${\cal Z}_3$ models}

$\bullet$
For an alternative way to define the parity operation we
 consider the model $A$ of the  ${\cal Z}_3$ monodromy discussed
 previously, (originally  given in~\cite{Dudas:2010zb}).
The massless spectrum is presented in Table~\ref{SU5Z3specA}.
   \begin{table}[t] \centering%
\begin{tabular}{|l|c|l|c|}
\hline
$SU(5)$& Equation &Matter&$R$-parity\\
\hline
$ 10^{(1)}_{t_i}=10_M$&$a_1$&$4Q+3u^c+5e^c$&$-$\\ \hline
$ \ov{10}^{(2)}_{-t_4}=\ov{10}_x$&$a_5$&$1\ov{Q}+2\bar e^c$&$-$\\ \hline
$ 5^{(0)}_{-2t_i}=5_{h_u}$&$a_1a_6a_8+\dots$&$h_u+D$&$+$\\ \hline
$ \bar 5^{(1)}_{t_i+t_4}=\bar 5_M$& $a_1a_6+a_2a_5$&$3(\ell+d^c)$&$-$\\ \hline
$  \bar 5^{(2)}_{t_{i}+t_5}$& $a_2a_7+a_1a_8$&$\ov{D}$&$-$\\ \hline
$ \bar 5^{(3)}_{t_{4}+t_5}$& $a_6a_7+a_5a_8$&$h_d$&$+$\\ \hline
\end{tabular}%
\caption{Field representation content under $SU(5)$, the defining equations and matter
content and parity
for the ${\cal Z}_3$ case (the indices $i,j,k$ take the values $1,2,3$).}
\label{SU5Z3specA}
\end{table}
 Following the same procedure
to each representation of the model we can associate a phase shown as in  the
second line of Table \ref{333Model}.
   \begin{table}[hbt!] \centering%
\begin{tabular}{l|lllllll}
\hline
Representation&$10^{(1)}$&$10^{(2)}$&$10^{(3)}$&$5^{(0)}$&$\bar 5^{(1)}$&$\bar 5^{(2)}$&$\bar 5^{(3)}$\\
phase\; $n\,\pi$&$\pi$&$5\,\pi$&$7\,\pi$&$15\,\pi$&$7\,\pi$&$9\,\pi$&$13\,\pi$\\
$e^{n{\rm mod}({13})i\pi}$&$-$&$-$&$-$&$+$&$-$&$-$&$+$\\
\hline
\end{tabular}%
\caption{ `Geometric' origin of Matter parity for the ${\cal Z}_3$ model A.}
\label{333Model}
\end{table}
The matter parity obtained by the operation (mod\,13) is given in the third
line of Table~\ref{333Model} and  is in accordance with
that one used in our preceding discussion of the ${\cal Z}_3$ model (see Table~\ref{SU5Z3specA}).
Although such an operation looks farfetched and a rather contrived attempt to match the
parities imposed by hand on the various models discussed, it still paves the way to consider
alternative -and possibly more realistic- methods to construct a parity consistent
with the local geometry.

$\bullet$
 Next, we discuss another case which associates
successfully matter parities to matter curves in a simpler  way.
Indeed, a rather simple assignment works for the  model $B$ of the ${\cal Z}_3$ case. Assume
that all $a_l$'s transform according to
\[ a_{l}\ra  (-)^{l+1}\,a_{l}\]

Using the defining equations (\ref{Z35split}) we find the following assignment
\[ 10^1(+),\; 5^0(+),\; \bar 5^1(-),\;\bar 5^2(-),\; \bar 5^3(-)\]
where the matter parity is shown in the brackets next to the representations.
This is compatible with the last column of Table~\ref{SU5Z3specB},  provided
we choose $\theta_{14}=(-),\; \theta_{15}=(+)$.

\subsection{ Matter parity for the ${\cal Z}_2$ model $C$.}

 For the model $C$ of the ${\cal Z}_2$ case, we can construct a parity
demanding that $a_i$'s transform as follows
\[ a_n \to a_n, {\rm for}\, n=6,7,\; a_m \to -a_m, {\rm for}\, m\ne 6,7.\]
We can check from (\ref{bksZ2}) that all $b_k$ transform the same way, $b_k\to -b_k$
and for $s\to s$ the spectral cover equation picks up only an overall minus sign.
From equations (\ref{10sZ2}) and (\ref{5spitZ2}) we observe that $5^{(0)}, 5^{(5)}$
and $10^{(3)}$ obtain a positive R-parity while the remaining representations
acquire negative R-parity. We observe that this matches the parity imposed on the model
of Table~\ref{restC} except for the $10^{(3)}$ representation which changes sign.
This would imply minor modifications of the model, the most important being on
the fermion mass textures. Indeed, recalling that  $10^{(3)}$
accommodates the lightest generation~\cite{Leontaris:2010zd}, we can see that by choosing the parities
of the singlets $\theta_{14}=(-)$ and $\theta_{43}=(+)$, we can generate
realistic fermion mass textures with affordable zero-entries.
Notice that the dangerous dimension four operators $10^{(3)}_M\bar 5^{(i)}_M 5^{(j)}_M\theta_{5k}$,
although now could be allowed, they  would require singlet vevs $\langle \theta_{5k}\rangle$
 which can be taken to be zero.

\subsection{A $Z_2$ matter parity for  a model with $2+3$ spectral cover split}

We now apply the idea of matter parity to a toy model with $2+3$ splitting.
This model was first  analyzed in refs~\cite{Marsano:2009wr,Dudas:2010zb},
however, we will see that in the present approach we are forced to
introduce a different R-parity.
The spectral cover equation is written as
\[ P_5=(a_1+a_2s+a_3s^2+a_4s^3)\,(a_5+a_6s+a_7s^2)\]

This leads to the identifications $t_{1,2,3}=t_a$ and $t_{4,5}=t_b$, with the
trace condition now reading $3t_a+2t_b=0$.
The equations connecting $b_k$'s with $a_i$'s are of the form $b_k\sim \sum_na_n a_{11-n-k}$
for appropriate values of $n$,  in particular
\be
\begin{split}
b_0&=a_4a_7\nn\\
b_1&=a_3a_7+a_4a_6\nn\\
b_2&=a_2a_7+a_3a_6+a_4a_5\\
b_3&=a_1a_7+a_2a_6+a_3a_5\\
b_4&=a_1a_6+a_2a_5\\
b_5&=a_1a_5\\
\end{split}
\ee
   \begin{table}[b] \centering%
\begin{tabular}{|l|c|c|l|c|c|c|}
\hline
$ SU(5)$&Equation&homology& $N_Y$&$M_X$&Matter&$R$\\
\hline
$ \ov{10}^{(1)}_{-t_a}=\ov{10}_x$&$a_1$ &$\eta-3c_1-\chi$&$-1$&$-1$&$(\bar Q,2\bar e^c)$&$-$\\ \hline
$ {10}^{(2)}_{t_b}=10_M$&$a_5$ &$-2c_1+\chi$&$1$&$4$&$(4 Q,3u^c,5e^c)$&$-$\\ \hline
$\bar 5^{(0)}_{t_a+t_b}=\bar 5_{h_d}$&$ a_1^2a_7+a_1a_2a_6+\cdots$ &$2\eta-6c_1-\chi$&$-1$&$0$&$h_d$&$-$\\ \hline
$  \bar 5^{(1)}_{2t_{a}}=5_M$& $a_1a_7+a_2a_6$ &$\eta-3c_1$&$0$&$-3$&$3(d^c,\ell)$&$+$\\ \hline
$  5^{(2)}_{-2t_{b}}=5_{h_u}$& $a_6$ &$-c_1+\chi$&$1$&$0$&$h_u$&$+$\\ \hline
\end{tabular}%
\caption{Field representation content under $SU(5)$, their homology class and flux
restrictions under $U(1)_{Y}$ for the ${\cal Z}_3\times {\cal Z}_2$ case (the indices
take the values $a=1,2,3$,  $b=4,5$).}
\label{23Model}
\end{table}
We further demand the following   transformations:
\begin{equation}
\label{500}
\begin{split}
a_m(\sigma(p))&= a_m(p)\, e^{im\pi}\;=\;(-)^m a_m(p),\;\;\;\;\;\;{\rm for}\; m=1,\dots, 7
\end{split}
\end{equation}

The constraint $b_1=a_3a_7+a_4a_6=0$ can be solved by $a_4=-\lambda\,a_7$ and $a_3=\lambda\, a_6$. Repeating
the steps as in the previous cases, we determine the homologies and flux restrictions of the
spectrum given in Table~\ref{23Model}.  Assuming the simplest scenario, we associate the matter parity
with the phase of the defining equation of the second column.  Using (\ref{500})
for $a_n$'s above, we obtain the parities of the last  column. Notice that
the latter does not coincide with the parity chosen in~\cite{Marsano:2009wr,Dudas:2010zb}.

Note, that there are  also $SU(5)$  singlets $\theta_{ij}$ obtained from the $24\in SU(5)_{\perp}$
residing on curves extended away from $S_{GUT}$. After the monodromy identifications they can be
organized  into  two categories. Those carrying $U(1)_i$-charges are  denoted with  $\theta_{ab}$,
 $\theta_{ba}$ and  the `neutral' ones $\theta_{aa},\theta_{bb}$ which can be identified with the
 neutrinos. Since we do not know the global geometry we will treat their parities as free parameters.

We distribute the matter and Higgs fields over the curves as follows
\[ 10_{t_b}=10_M,\; \bar 5_{t_a+t_b}=\bar 5_{h_d},\; 5_{-2t_b}=5_{h_u},\;
 \bar 5_{2t_a}=\bar  5_M,\;\ov{10}_{-2t_a}=\ov{10}\]
 while we determine the multiplicities by choosing $N=1$ and the $M_X$'s as shown in tha
 Table~\ref{23Model}.

The  allowed tree-level couplings with non-trivial $SU(5)$ representations are
\be
\begin{split}
{\cal W}_{tree}&= \lambda_u\,10_M\,10_M\,\bar 5_{h_u}+\lambda_d\,10_M\,\bar 5_M\,\bar 5_{h_d}
+ \lambda_x\,\ov{10}_x\,10_M\theta_{ab}
\end{split}
\ee
The first two  provide masses to quarks and charged leptons. The third term
survives by assuming  positive matter parity for $\theta_{ab}$, while
a non-zero vev gives masses to the exotic matter.

The dangerous dimension four and five operators (namely $10_M\bar 5_M\bar 5_M$ and $10_M 10_M 10_M\bar 5_M$)
inducing proton decay are eliminated under the combined action of parity and $U(1)$ symmetry.
Notice also that the Higgs mixing term $5_{h_u}\bar 5_{h_d}\theta_{ab}$
is prevented by R-parity.

\newpage

\section{Conclusions}

In the present work we have analyzed several aspects of the local F-theory  GUTs
associated to  $SO(10)$ and $SU(5)$ singularities of the internal geometry.
We have considered the analysis in  a spectral cover  context  where
these symmetries are incorporated in  $E_8$
which is assumed to be the maximum singularity of the internal manifold.
We have investigated several implications on the derived models of all possible monodromies
among the $U(1)$ symmetries  emerging from the  $SU(4)_{\perp}$ and the $SU(5)_{\perp}$
spectral covers  corresponding to $SO(10)$ and $SU(5)$ gauge symmetries respectively.

In particular, we have investigated systematically the landscape of effective
models with $SO(10)$ and $SU(5)$ gauge symmetries emerging  under the various
cases of monodromies among abelian factors embedded in the $SU(4)_{\perp}$ and
$SU(5)_{\perp}$ spectral  cover respectively.  Moreover, we have examined all
possible ways that the abelian  factors undergoing monodromies are embedded in
the enhanced symmetries at the points of double and triple
intersections of seven branes.
 We have explored the implications  of  the successive $U(1)$-flux breaking of
the $SO(10)$ and $SU(5)$ gauge symmetries down to the Standard Model gauge group.
Using the combined data of the $U(1)_X$ flux breaking $SO(10)\to SU(5)\times U(1)_X$
and the $U(1)_Y$ hypercharge flux breaking of $SU(5)$ GUT down to SM,  we have determined
 the  induced restrictions on the multiplicities of the massless spectrum of the effective
field theory models. Following the  described procedure, we have  built several examples
of models and discussed their viability as well as their low energy  massless spectrum.
Despite the combined constraints arising form the GUT symmetries, monodromies and
fluxes, we have seen that  not all dangerous baryon couplings are eliminated, unless
a matter parity is associated to the various massless states. We have considered
the possibility that a discrete matter parity  emanating from the geometric
properties of the internal manifold can be communicated to the matter curves.
We have confirmed the successful implementation of this geometric concept of
matter parity in  several examples constructed within the proposed scenario
of this paper.  Further investigations on the specific properties
 of these models would be required to discriminate them with respect to
their low energy implications, however these are beyond the scope
of the present work. We plan to revisit these interesting issues
in a future publication.

\vfill

\section*{Acknowledgements}
 This work was supported in part by the European Commission
under the ERC Advanced Grant 226371 and the contract PITN-GA-2009-237920.
 GKL would like to thank CERN, Theory Division for kind hospitality where part of these
work has been carried out.


\newpage

\appendix{{\Large\bf{Appendix}}}

\section{F-$SO(10)$ gauge symmetry and enhancements}

In this appendix we summarize some useful formulae and present the basic techniques
for model building through the spectral cover approach.  Furthermore, we examine
in some detail the monodromic $U(1)$'s accompanying the $SO(10)$ and $SU(5)$ gauge
symmetries.

 We have seen that the  $SO(10)$  model  has  a spectral cover characterized by a
 $SU(4)_{\perp}$ symmetry. We  assume a single point of  ${\cal E}_8$ enhanced symmetry with chiral
matter and Higgs descending from the ${\cal E}_8$ -adjoint representation. Therefore,  we start with the
decomposition of the adjoint~(\ref{E8adj}) under the breaking pattern (\ref{E82U4_1}).
To obtain the effective $SO(10)$ model, we  further assume the breaking of $SU(4)\ra U(1)^3$ by flux effects. Each of the $SO(10)$ representations lies on a matter curve which is distinguished by the specific
charge it carries under the  $U(1)^3$ Cartan subalgebra,
characterized by the $SU(4)_{\perp}$ weights denoted with $t_i, i=1,2,3,4$. As it happens for any $SU(N)$ symmetry they satisfy the tracelessness condition
\[t_1+t_2+t_3+t_4=0\]
  There are in principle four matter curves (denoted with $\Sigma_{16}$) accommodating the ${\bf 16}$ representations and an equal number of ${\bf \ov{16}}$ matter curves, six  $\Sigma_{10}$ matter curves for the {\bf 10}, and fifteen singlets $\Sigma_{1}$.
   The $SU(4)_{\perp}$ weights distinguishing the ${\bf 16}$'s, ${\bf 10}$'s and twelve singlets are given
 in  (\ref{specurves}).
Then, the matter fields are localized on the curves which lie in the following directions
of the Cartan subalgebra
\[{\bf 16}:\, t_i=0,\; {\bf 10}: \, t_i+t_j=0,\; {\bf 1}: \, t_i-t_j=0\]
In $SO(10)$ the Yukawa coupling giving mass to fermion fields is ${\bf 16\, 16\, 10}$. Since
the $SO(10)$ representations carry  $U(1)_i$ charges, this coupling should also be invariant
under these abelian factors
\[{\bf 16}_{t_i}\, {\bf 16}_{t_j}\, {\bf 10}_{t_k+t_l}\]
$U(1)$ invariance is ensured by the condition $t_i+t_j+t_k+t_l=0$.  Since  these indices span the numbers $1,2,3,4$, this automatically implies that all indices $i,j,k,l$ differ from each other,
so that we get
\[t_i+t_j+t_k+t_l=t_1+t_2+t_3+t_4=0\]
Thus, invariance under  $U(1)_i\in SU(4)_{\perp}$  would require the ${\bf 16}$'s to descend from different matter curves. For example, if Higgs doublets are found in  ${\bf 10}_{t_3+t_4}$,
  then the only available couplings arise from ${\bf 16}_{t_1} {\bf 16}_{t_2}{\bf 10}_{t_3+t_4}$,
  leading to off-diagonal tree level masses involving at least two generations.
  The known hierarchical fermion mass spectrum and the heaviness of the
  third generation however, is compatible with rank one structure of the
mass matrices at tree-level.
 This requires a solution where at least two of the curves are identified through
some (discrete) symmetry which has to be a subgroup of the Weyl group $W(SU(4))=S_4$.
For example, assuming the simplest case, namely a ${\cal Z}_2$ symmetry among
$t_1 \leftrightarrow t_2$, we obtain the identification ${\bf 16}_{t_1}={\bf 16}_{t_2}$.
In this case we interpret ${\bf 16}_{t_1} {\bf 16}_{t_2}{\bf 10}_{t_3+t_4}$ as a diagonal Yukawa coupling which provides masses to the third generation fermion fields.

\subsection{Review of the Weierstrass form and some related material}

According to the `standard' interpretation  in F-theory the  gauge symmetry is associated
to the singularities of the internal compact manifold.  A systematic  analysis of these
singularities has started with the work of Kodaira.  Given the form of the Weierstrass equation
\[ y^2=x^3+f(z)x+g(z)\]
the Kodaira classification relies on the vanishing order of the
polynomials $f,g$ and the discriminant $\Delta$.  This is summarized
in Table~\ref{T_1}.
\begin{table}
\begin{center}
\begin{tabular}{|l|l|l|l|l|}
ord($f$)& ord($g$)& ord($\Delta$) &fiber type &Singularity\\
\hline
$0$&$0$&$n$&$I_n$&$A_{n-1}$
\\
$\ge 1$&$1$&$2$&$II$&none
\\
$1$&$\ge 2$&$3$&$III$&$A_1$
\\
$\ge 2$&$2$&$4$&$IV$&$A_2$
\\
2&$\ge 3$&$n+6$&$I_n^*$&$D_{n+4}$
\\
$\ge 2$&$ 3$&$n+6$&$I_n^*$&$D_{n+4}$
\\
$\ge 3$&4&8&$IV^*$&$E_6$
\\
$ 3$&$\ge 5$&9&$III^*$&$E_7$
\\
$\ge 4$&5&10&$II^*$&$E_8$
\\
\hline
\end{tabular}
\caption{Kodaira's classification of Elliptic Singularities. The first three
columns refer to the order of vanishing of $f,g,\Delta$ polynomials with respect to $z$.
Column four denotes the type of the fiber ($I$ nodal, $II$ cuspidal etc). Column
five designates the associated singularity. }\label{T_1}
\end{center}
\end{table}
 A useful tool for the analysis of the gauge properties
of an F-theory GUT is Tate's algorithm.
Tate's Algorithm~\cite{Tate75} provides a method to describe  the singularities
of the elliptic fiber and determine the local properties of the associated  gauge
group.
\begin{table}[!t]
\centering
\renewcommand{\arraystretch}{1.2}
\begin{tabular}{|c|c|c|c|c|c|c|c|}
\hline
Type &{\bf Group} & ${ a_1 }$& ${ a_2 }$& ${ a_3} $& ${ a_4 }$& ${a_6} $& ${ \Delta}$\\
\hline
$I_0$& ${0}$&0&0&$0$&$0$&$0$&$0$   \\
\hline
$I_1$& $-$&0&0&$1$&$1$&$1$&$1$   \\
 \hline
$I_2$&  $-$&0&0&$1$&$1$&$2$&$2$   \\
 \hline
$I_{2n}^s$& ${ SU(2n)}$&0&1&$n$&$n$&$2n$&$2n$   \\
\hline
$I_{2n+1}^s$& ${ SU(2n+1)}$&0&1&$n$&$n+1$&$2n+1$&$2n+1$   \\
 \hline
$I_1^{*s}$& ${ SO(10)}$&1&1&$2$&$3$&$5$&$7$   \\
\hline
$I_{2k-3}^{*s}$& ${ SO(4k+2)}$&1&1&$k$&$k+1$&$2k+1$&$2k+3$   \\
 \hline
$IV^{*s}$& ${E_6}$&1&2&$3$&$3$&$5$&$8$   \\
 \hline
$III^{*s}$& ${ E_7}$&1&2&$3$&$3$&$5$&$9$   \\
 \hline
$II^{s}$& ${ E_8}$&1&2&$3$&$4$&$5$&$10$  \\
 \hline
\end{tabular}
 \caption{ Partial results of  Tate's Algorithm. (The complete results can be found in~\cite{Bershadsky:1996nh}). The order of vanishing of the coefficients ${ a_i\sim z^{n_i}}$ and the corresponding {gauge group}. The highest singularity allowed in the elliptic fibration is ${ E_8}$.\label{dpn} }
\end{table}

To study the semi-local model, we need to determine the properties of the matter
curves and in particular how the elliptic fibration degenerates on the GUT surface
$S_{GUT}$. For local analysis, a suitable form of Weierstrass equation (Tate's form) is
   \ba
 {y^2}+{ a_1}{ x\,y}+{ a_3}{ y}&=&
 { x^3}+{a_2}\,{ x^2}+{ a_4}{ x}+{ a_6}\label{Wei}
   \ea
   with $a_n$ being polynomial functions on the base.
The indices of the coefficients $a_n$ have been chosen so to indicate the section they
belong to, i.e. $a_n\in K_{B_3}^{-n}$. Thus  each term is a section $K^{-6}_{B_3}$
(see \cite{Donagi:2008ca} for details.).

The standard form of the Weierstrass equation is
\[y^2=x^3+fx+g\]
and is obtained by completing the square and the cube, as
follows. The square on the left hand side becomes
\ba
\left(y+\frac{a_1x+a_3}{2}\right)^2&=& { x^3}+{a_2}\,{ x^2}+{ a_4}{ x}+{
a_6}+\left(\frac{a_1x+a_3}{2}\right)^2\nn
\ea
while we equate the RHS with
\[\left(x+\lambda\right)^3+f\,(x+\lambda)+g\]
Comparing, we get
\ba
f&=&\frac{1}{48} \left(24 \,a_1 \,a_3-\left(a_1^2+4 \,a_2\right)^2\right)+ a_4\nn\\
g&=&\frac{1}{864} \left(a_1^6+12 a_1^4 a_2-36 a_1^3 a_3+48 a_1^2 a_2^2\right.\nn\\&&\left.
-72 a_4 \left(a_1^2+4
   a_2\right)-144 a_1 a_2 a_3+64 a_2^3+216 a_3^2\right)+a_6
\ea
Using the definitions
\[
\b_2=a_1^2+4 a_2,\;\b_4=a_1a_3+2a_4,\;\b_6=a_3^2+4 a_6\]
the functions $f,g$ can be rewritten in a simpler form
\ba
f&=&-\frac{1}{48}\left(\b_2^2-24 \b_4\right)\nn\\
g&=&-\frac{1}{864}\left(-\b_2^3+36\b_2\b_4-216\b_6\right)
\ea
If we further define
\[\b_8=\b_2a_6-a_1a_3a_4+a_2a_3^2-a_4^2\]
we can write the discriminant
\be
\label{discr}
\begin{split}
\Delta&=4f^3+24g^2\\
      &=-\b_2^2\b_8-8\b_4^3-27\b_6^2+9\b_2\b_4\b_6
\end{split}
\ee
$f,g$ are assumed to be functions of a complex coordinate $z$ on the base $B_3$.

We can now associate the vanishing order of $f(z), g(z), \Delta(z)$  to the singularity
type of the compact manifold as in Table~\ref{T_1}. On the other hand, Table~\ref{dpn}
associates the coefficients $a_n$ to the singularity.

\subsection{$SO(10)$}

We now apply the above analysis to the case of interest, namely the $SO(10)$ model.
 Using the Tate's algorithm~\cite{Tate75}, for the $SO(10)$ we substitute the coefficients $b_i$
 with
 \[a_1=-b_5 z,\;a_2=b_4\,z,\;a_3=-b_3z^2,\;a_4=b_2z^3,\;a_6=b_0z^5\]
 and we can write the Weierstrass equation as follows
\be
\label{Wei4}
\begin{split}
y^2&=x^3+b_5xyz+b_4x^2z+b_3yz^2+b_2xz^3+b_0z^5
\end{split}
\ee
We mention that $x,y$ are homogeneous coordinates of the torus fiber and $b_i$
functions of the coordinates of the three-fold base. The coefficients $b_i$
are non-vanishing and may have subleading terms being powers of $z$.
 We can study the discriminant and determine the singularity enhancements along
 the lines of  ref~\cite{Donagi:2008ca}.

For the local picture (i.e. in the limit $z=0$) we recall that
the $b_i$-subleading terms vanish and $b_i$ become constants
that can be interpreted as sections of line bundles on the surface $S_{GUT}$.
 The discriminant is
 \ba
  \Delta&=&-(16 b_3^2 b_4^3)\, z^7\nn
 \\
 &+&\left(27 b_3^4-36 b_4 b_5 b_3^3+8 b_4 \left(b_4 b_5^2-9 b_2\right) b_3^2-16 b_2 b_4^2 b_5 b_3+16 b_4^2
   \left(4 b_0 b_4-b_2^2\right) \right)z^8\nn
   \\&+&  \left[64 b_2^3-8 b_5 \left(b_4 b_5-12 b_3\right) b_2^2-2 \left(b_3 \left(4 b_4 b_5-15 b_3\right)
   b_5^2+144 b_0 b_4\right) b_2\right.\nn\\
   &+&\left. b_3^2 b_5^3 \left(b_4 b_5-b_3\right)+24 b_0 \left(9 b_3^2-6 b_4
  b_5 b_3+2 b_4^2 b_5^2\right)\right]\,z^9+\cdots
\ea
while the functions $f,g$ are
\ba
f(z)&=&\frac{1}{48} \left(24 \left(2 b_2 +b_3 b_5\right) z^3-\left(z^2 b_5^2+4 z
   b_4\right){}^2\right)\nn\\
   g(z)&=&
   \frac{1}{864} \left(\left(z^2 b_5^2+4 z b_4\right){}^3-36 \left(2 b_2 z^3+b_3 b_5 z^3\right)
   \left(z^2 b_5^2+4 z b_4\right)+216 \left(4 b_0 z^5+b_3^2 z^4\right)\right)\nn
   \ea
\subsubsection{Symmetry enhancement}
From the above formulae we see that the coefficients in the lowest powers of $z$ are
\be
\begin{split}
f(z)&=-\frac{b_4^2}{3}\,z^2+\frac{1}{48} \left(-8 b_4 b_5^2+24 b_3 b_5+48 b_2\right)\,z^3+{\cal O}(z^4),\\
 g(z)&=\frac{2 b_4^3}{27}\,z^3+{\cal O}(z^4)
 \end{split}
 \ee
  Clearly, we can see from Table~\ref{T_1} that  this is indeed a $D_5=SO(10)$ singularity
  since they satisfy deg$\,f= 2$, deg$\,g=3$ and deg$\,\Delta =7$. There are several ways to
  enhance this symmetry:

$i$) Setting $b_4=0$, to lowest order in $z$ we get
\[f(z)=\left(b_2+\frac{b_3 b_5}{2}\right)\,z^3+{\cal O}(z^4),\;g(z)=\frac{b_3^2}{4}\,z^4+{\cal O}(z^5),\;
\Delta(z)=b_3^4\,z^8+{\cal O}(z^9)\]
According to Kodaira's classification along this intersection we get  an ${\cal E}_6$
 enhancement of the $SO(10)$ singularity where the ${\bf 16}$ of $SO(10)$ resides.
 This can be seen from the decomposition
\[27\ra {\bf 16}_1+{\bf 10}_{-2}+{\bf 1}_4\label{27}\]

Next, we consider  a second order enhancement with the case when both $b_3=b_4=0$.
We obtain
\[f(z)=b_2\,z^3+{\cal O}(z^4),\;g(z)=\left(b_0-\frac{1}{12} b_2 b_5^2\right)\,z^5+{\cal O}(z^6),\;
\Delta(z)=-64 b_2^3\,z^9+{\cal O}(z^{10})\]
Thus, this corresponds to an $E_7$ enhancement. Decomposition of the $E_7$ representations
\[56\ra 27+\ov{27}+1+1,\; 133\ra 78+27+\ov{27}+1\]
entails the realization of the $SO(10)$ Yukawa coupling ${\bf 16\,16\,10}$,
since
\[56\cdot 56\cdot 133\ra 27\cdot 27\cdot (27+ 78)\ra 16_1\cdot 16_1\cdot 10_{-2}\]

$ii$) We may study other types of enhancements as follows. Consider first that
$b_3=0$. Then, to lowest order in $z$, we have
\[f(z)=-\frac{b_4^2}{3}\,z^2+{\cal O}(z^3),\;g(z)=\frac{2 b_4^3}{27}\,z^3+{\cal O}(z^4),\;
\Delta(z)=16 b_4^2 \left(b_2^2-4 b_0 b_4\right)\,z^8+{\cal O}(z^9)\]
which corresponds to the enhancement $D_6=SO(12)$.
We have the following decompositions along this singularity enhancement
\[66\ra 45+1+10_2+\ov{10}_2,\; 32\ra 16_1+\ov{16}_{-1},\; 12\ra 10_0+1_2+1_{-2}\]
Thus Higgs fields are found in the decomposition of $66$ adjoint while half of $32$
representation corresponds to ${\bf 16}$.

 This singularity is further enhanced to $D_7=SO(14)$ if in addition to $b_3=0$ we also impose
\ba
 b_2^2-4 b_0 b_4=0&\ra & {\rm deg}\Delta =9\label{so12en}
 \ea
The $SO(12)$ representations are found in the decompositions
\[14\ra 12+1+1,\; 91\ra 66+1+12+12'\]

Now, let us examine how this singularity looks like locally. A way to
 obtain a local model from a global one, is to assign scaling dimensions to
$(x, y, z)$ and retain only the relevant terms. To this end,
 we introduce the scaling dimensions~\cite{Donagi:2008ca,Chen:2010ts}
 $(x,y,z)\sim (\frac 13,\frac 12, \frac 15)$
 and observe that the terms of order one recreate the $E_8$ singularity
 \[ y^2 =x^3+b_0z^5\]
The term $b_5xyz$  is of order higher than one while all  the remaining all less than one.
Dropping the term with scaling greater than one we obtain the local deformation of
the $E_8$ singularity
\ba
y^2&=&x^3+b_4x^2z+b_3yz^2+b_2xz^3+b_0z^5\label{so10e}
\ea
which does not depend on $b_5$.


We can see how this is `encoded' locally into an $SU(4)$ spectral cover along the intersection
curve by identifying $z^5$ with the fourth power of an affine parameter  $s$
\[ z^5 \equiv s^4\; \ra z=s^{4/5}\]
Let now fix the $x$ scaling with respect to $s$ by demanding that the term of $b_2$ coefficient
is a power of $s^2$
\[xz^3 =s^2\ra x= s^2z^{-3}= s^{-2/5}\]
We further require that the  $b_3$ coefficient multiplies the first power of  $s$, thus
\[yz^2=s\ra y=s z^{-2}=s^{-3/5}\]
All remaining terms are now fixed. In particular we find also that $y^2=s^{-6/5}=x^3$.
 In the spectral cover equation (\ref{so10e})  now becomes
\ba
0&=&b_4+b_3s+b_2s^2+b_0s^4
\ea
This is indeed an $SU(4)$ spectral cover with the $b_i$ coefficients  as in (\ref{so10e}).
($b_1=0$  as expected for any $SU(N)$).

\subsection{Polynomial equations for the matter curves}

We can proceed with the analysis of the $SO(10)$ GUT models  using the equivalent
description of  the spectral cover approach.   More precisely  we can describe the model
in the context of the Higgs bundle picture which is given in terms of the adjoint scalars
and the gauge field.

 We have seen that for the $E_8$ embedding of the $SO(10)$ singularity  the commutant
 is  $SU(4)$, while  this is given by the hypersurface
\ba
{\cal C}_4=\sum_{k=1}^4b_ks^{4-k}&=&b_0s^4+b_1s^3+b_2s^2+b_3s+b_4=0
\label{4sc}
\ea
with $s$ being an affine parameter and $b_1=0$. This is the spectral cover
for the fundamental representation of $ SU(4)$.
We denote with $c_1$ the $1^{st}$  Chern class of the   Tangent Bundle to $S_{GUT}$
and $-t$ the $1^{st}$ Chern class of the Normal Bundle to $S_{GUT}$.
It is customary  to define the following  quantity
\[\eta =6\,c_1-t\]
Using this, we can express the coefficients $b_k,\; k=0,\dots,4$ as sections of
\ba
[b_k]=\eta-k\,c_1=(6-k)c_1-t\label{secc}
\ea
while $[s]=-c_1$, so that each term in (\ref{4sc}) is $[b_ks^{4-k}]=\eta-4c_1$.

 We can determine the `locations' $t_i$ of the four
 ${\bf 16}$ representations as the roots of the  polynomial
\ba
P_{16}(s)&=&b_0\left(s-t_1\right) \left(s-t_2\right) \left(s-t_3\right) \left(s-t_4\right)\nn\\
  &=&b_0s^4+b_1s^3+b_2s^2+b_3s+b_4\label{16eq}
\ea
We can identify the parameter $s$ with the Higgs vev  breaking ${\cal E}_8$.
Setting $s=0$, we can see that the equation
\[P_{16}(s=0)=0\,\Rightarrow\,  b_4=t_1t_2t_3t_4=0\]
determines the `locations' of the four matter curves which lift to a single one
in the spectral cover.

It is useful to derive the  equations relating coefficients $b_k$ and  $t_i$. These are
\ba
b_1&=&-b_0 (t_1+t_2+t_3+t_4)=0\nn\\
b_2&=&b_0(t_1^2+t_2^2+t_3^2+ t_1t_2+t_2 t_3+t_3 t_1)\nn\\
b_3&=&b_0\left(t_1+t_2\right) \left(t_1+t_3\right) \left(t_2+t_3\right)\nn
\\
b_4&=&b_0t_1t_2t_3t_4=-b_0t_1t_2t_3(t_1+t_2+t_3)\label{bts}
\ea
where the solution of  $b_1=0\ra t_4=-(t_1+t_2+t_3)$  has been substituted into $b_{2,3,4}$.

Next we construct the spectral cover for the antisymmetric representation ${\bf 10}\in SO(10)$.
The  ${\bf 10}$ representations are characterized by the weights $t_i+t_j$ with
$i,j=1,2,3,4$.
Proceeding as in the case of the ${\bf 16}$'s,
we can write the equation for the ${\bf 10}$'s of $SO(10)$ as follows:
\ba
P_{10}(s)&=&b_0^2\prod_{i<j}(s+t_i+t_j)\nn\\
&=&b_0^2\left(s-t_1-t_2\right) \left(s+t_1+t_2\right) \left(s-t_1-t_3\right) \left(s+t_1+t_3\right)\left(s-t_2-t_3\right)
   \left(s+t_2+t_3\right)\nn\\
   &=& b_0^2s^6+c_1s^5+c_2s^4+s_3s^3+c_4s^2+c_5s+c_6\label{10eq1}
\ea
For later use, we  express $c_n$ in terms of $t_i$ and then using equations (\ref{bts})
we convert them to  functions of $b_k$
\be
\label{gcMU}
\begin{split}
c_2&=-2b_0^2 (t_1^2+t_2^2+t_3^2+ t_1t_2+t_2 t_3+t_3 t_1)b_0^2=-2b_2b_0^2\\
c_6&=-b_0^2\left(t_1+t_2\right){}^2 \left(t_1+t_3\right){}^2 \left(t_2+t_3\right){}^2=-b_3^2\\
{c_4}&={b_0^2}\left(t_1^4+2 \left(t_2+t_3\right) t_1^3+\left(3 t_2^2+8 t_3 t_2+3 t_3^2\right) t_1^2\right.\\
&\;\;\;\;\;\;\;\;\;\;\left.+2 \left(t_2+t_3\right) \left(t_2^2+3
   t_3 t_2+t_3^2\right) t_1+\left(t_2^2+t_3 t_2+t_3^2\right){}^2\right)
\end{split}
\ee
and the last one can be rewritten
\[c_4=b_2^2+4b_0^2 t_1 t_2 t_3 \left(t_1+t_2+t_3\right)\equiv b_2^2-4b_4b_0\]
Thus, all coefficients are in terms of $b_i$'s and $P_{10}$ takes the simple form
\ba
P_{10}(s)&=&b_0^2s^6-2b_2b_0\,s^4 +(b_2^2-4b_4b_0)s^2-b_3^2\label{10eqA}\ea
Setting $s=0$, we see then that the 7-branes  associated to ${\bf 10}$'s of $SO(10)$
are determined by
\[b_3^2=0\,\cdot\]

\begin{table}
\begin{center}
\begin{tabular}{|l|l|l|}
\hline
Order&Equation&Enhancement \\
\hline
$1^{st}$&$b_4=0$&$E_6$
\\
$2^{nd}$&$b_4=b_3=0$&$E_7$
\\
\hline
$1^{st}$&$b_3=0$&$SO(12)$
\\
$2^{nd}$&$b_3=b_2^2-4b_0b_4=0$&$SO(14)$
\\
\hline
\end{tabular}
\caption{First and second order enhancements of $SO(10)$. The $b_i$ are
the coefficients of the corresponding Weierstrass equation (\ref{Wei4}). }
\end{center}
\label{T_e}
\end{table}

\section{Monodromies}

Much  of the F-theory edifice  rests on the notion of monodromies.
Matter curves are associated to the roots $t_i$ which are
polynomial solutions with factors combinations of  $b_i$'s, thus
\[ b_i=b_i(t_j)\]
Generically, the inversion of these equations will lead to branchcuts and
he solutions $t_j=t_j(b_i)$ are subject to monodromy actions.
There are several ways to factorize  the spectral cover equation,
the most obvious possibilities
are $2+1+1, 2+2$ and $3+1$ corresponding to  (\ref{z2case_1},\ref{z2z2case_1}) and (\ref{z3case_1})
respectively.
For the first case we have a single ${\cal Z}_2$ monodromy among $\{t_1,t_2\}$. In the second case
we get the identifications between $\{t_1,t_2\}$ and similarly among $\{t_3,t_4\}$ implying a
${\cal Z}_2\times {\cal Z}_2$
monodromy and finally in case 3 we have a ${\cal Z}_3$ monodromy among $\{t_1,t_2,t_3\}$.  Next
we analyze in detail these three cases.

\subsection{Symmetry enhancements and ${\cal Z}_2$ monodromy}

 The spectral cover equation for the factorization  $C_{2+1+1}$ is given in (\ref{z2case_1})
and  corresponds to a ${\cal Z}_2$ monodromy among $t_1\leftrightarrow t_2$.  Putting $s=0$  we find
\[P_{16}(0)=a_1a_4a_6=0\]
thus, there are three  ${\bf 16}$'s left after the monodromy action which are in
\[a_1=0,\;a_4=0,\;a_6=0\]
The determination of the $a_i$ homologies  can be achieved through their
 link to the known $b_k$'s. Comparing powers of $s$ between (\ref{4sc}) and (\ref{z2case_1})
we get
\be
\label{Z2bas}
\begin{split}
b_4&=a_1 a_4 a_6\\
b_3&=a_2 a_4 a_6+a_1 a_5 a_6+a_1 a_4 a_7\\
b_2&=a_3 a_4 a_6+a_2 a_5 a_6+a_2 a_4 a_7+a_1 a_5 a_7\\
b_1&=a_3 a_5 a_6+a_3 a_4 a_7+a_2 a_5 a_7\\
b_0&=a_3 a_5 a_7
\end{split}
\ee
  We solve the constraint $b_1=0$ adopting the following Ansatz:
\be
\label{Ansatz}
\begin{split}
 a_3&=\lambda a_5 a_7,\;\; a_2=-\lambda (a_5a_6+a_4a_7)
 \end{split}
 \ee
Given that  the coefficients $b_k$
satisfy (\ref{secc}), for the specific combinations
of indices appearing in (\ref{Z2bas}) we have
\[\eta-k\,c_1=[a_l]+[a_m]+[a_{n}],\;\; {\rm with} \, k+l+m+n=15
\]
where $l,m,n$ take the values $1,2,\dots, 7$ and $k=0,1,2,3,4$.
These are five equations with seven unknowns. We choose
two arbitrary values $[a_5]=\chi_5, [a_7]=\chi_7$ while for
convenience  we introduce
\[\chi=\chi_5+\chi_7\]
 and solve the  system. The results are presented in Table~\ref{T_0A}.
\begin{table}
\begin{center}
\begin{tabular}{c|c|c|c|c|c|c}
\hline
$a_1$& $a_2$& $a_3$& $a_4$& $a_5$& $a_6$& $a_7$
\\
$\eta-2c_1-\chi$& $\eta-c_1-\chi$& $\eta-\chi$& $-c_1+x_5$& $x_5$& $-c_1+x_7$& $x_7$\\
\hline
\end{tabular}
\end{center}
\caption{Homology classes for coefficients $a_i$  for the ${\cal Z}_2$ case}
\label{T_0A}
\end{table}
The homology of $\lambda$ can be specified using the homologies of $a_i$  presented in Table~\ref{T_0A}:
\[[\lambda]=\eta-2\chi =6c_1-t-2\chi\]
\begin{table}
\begin{center}
\begin{tabular}{l|l|l|l}
\hline
${\bf 10}$&$(a_1-\lambda a_4a_6)$&$\eta-2c_1-\chi$&$t_{1,2}+t_3/t_4$
\\
${\bf 10}$&$(a_5a_6+a_4a_7)$&$-c_1+\chi$&$t_{3}+t_4$\\
\hline
\end{tabular}
\end{center}
\caption{Homology classes of ${\bf 10}$'s  for the ${\cal Z}_2$ case}
\label{T1x}
\end{table}
Substituting the solution for $b_1=0$, the $b_k$ coefficients become
\ba
b_0&=&\lambda\, (a_5a_7)^2\nn\\
b_1&=&0\nn\\
b_2&=&a_5a_7 (a_1+\lambda\,a_4a_6)-\lambda  (a_5a_6+a_4a_7)^2\label{asol}\\
b_3&=&(a_1-\lambda a_4a_6)(a_5a_6+a_4a_7)\nn\\
b_4&=&a_1a_4a_6\nn
\ea
We have already investigated the coefficients $c_i(b_k)$ of the polynomial associated to $\Sigma_{10}$
matter curves  and found that their equations satisfy $b_3^2=0$.  Therefore their homologies are specified
by the homologies of the factors constituting $b_3$. Notice that the coefficient $b_3$  appearing in the
above solution  is already factorized, its factors written as simple combinations of $a_i$. Since all $[a_i]$
classes are specified, it is straightforward to see that equation $b_3^2=0$ defines four ${\bf 10}$-matter
curves with homology classes determined straightforwardly from those of $a_i$'s and given in Table~\ref{T1x}.
 On the other hand we know that under the $t_1 \leftrightarrow t_2$ identification the six ${\bf 10}$'s
 characterized by the  $t_i+t_j$ reduce to four, in accordance with the factorization of $b_3$.
We collect all the results in Table~\ref{T_ALL}.

\subsubsection{Enhancements along intersecting matter curves}


We have seen previously how the various singularity enhancements along matter curves
 are attributed to  the vanishing of some $b_i$'s. Here we will  associate
 these enhancements directly to the coefficients $a_i$.  We note that this analysis can help us
 specify exactly  which $U(1)$'s are embedded  in the enhanced gauge group.  Further, since
it happens that the coefficients $a_i$ are directly related to monodromies,
 this  might be useful in  phenomenological applications. For example, the computation
 of the Yukawa couplings requires knowledge of the wavefunctions of the states participating
 at the triple intersection.  The wavefunction profiles are determined by the solution of
  a set of  differential equations obtained from varying the equations of motion~\cite{Beasley:2008dc}.
  The computation of the wavefunctions and the Yukawa coupling depend crucially on the whether
  $U(1)$ in the intersection undergo monodromies or not~\cite{Cecotti:2010bp}.

{\bf  $E_6$  and $E_7$ Enhancement}

We know already that  this enhancement is obtained setting $b_4=0$. At the level of
the $a_i$ coefficients this can be done
by demanding either of $a_{1,4,6}$ to be zero. There is a difference however
between  $a_{1}=0$ and  $a_{4,6}=0$. The first case (i.e. $a_1=0$ which is involved
in the two $t_{1,2}$ undergoing a monodromy) implies
\[ s a_2+s^2 a_3=s (a_2+a_3 s) \]
This means that for the particular choice $a_1=0$ at the first order enhancement
to $E_6$  the monodromy is `resolved'  in the sense that  one of the two $U(1)$ involved
($s=0$) is incorporated to the $E_6$ symmetry.

 Here, we make the  general observation that if  the monodromy is among $U(1)$'s (or the $t_a$'s) which are
 embedded into the $E_6$ enhancement,
 at this enhanced symmetry level we have distinct $27_{t_{1,2,3}}$ matter curves and the corresponding wavefunctions
refer to different entities at this stage.  On the contrary,
if  the monodromy is in $U(1)_i\in SU(3)_{\perp}$ then some $27_{t_i}$ are identified.
We reckon that these distinct cases might have some relevance on  the determination
of the wavefunctions of the states participating in the vertex and as
a consequence to the Yukawa coupling computations~\cite{Cecotti:2010bp}, however we leave
such an analysis for a future work.

For  the other two cases ($a_{4,6}=0$) the monodromy among the two $U(1)$'s
is preserved at the first order enhancement of the symmetry.

$\bullet$  The condition $b_4=0$ is satisfied by setting any of $a_1,a_4,a_6$ zero.
 Thus, let $a_1=0$. Checking the discriminant and the vanishing
order of the coefficients in Weierstrass equation, we find
\ba
f(z)&=&\left(\lambda a_4 a_5 a_6 a_7-\lambda\left(a_5 a_6+ a_4 a_7\right)^2\right)z^3+\cdots,\nn
\\
g(z)&=&\frac{1}{4} \lambda^2 a_4^2 a_6^2 \left(a_5 a_6+ a_4 a_7\right){}^2 z^4+\cdots,\nn
\\
\Delta(z)&=&-27 \lambda^4 a_4^4 a_6^4 \left(a_5 a_6+ a_4 a_7\right){}^4z^8+\cdots\nn
\ea
that is, we obtain an $E_6$ enhancement.

In addition we set now $b_3=0$, which, as we have seen enhances the symmetry to $E_7$.
With respect to $a_i$, this happens when either of the following occurs:
\[ a_{4}=0,\; a_6=0,\;  a_5 a_6+ a_4 a_7=0\]
$a$) The first two cases are equivalent. Accepting $a_4=0$ the spectral cover equation becomes
\[{\cal C}_4= s^2(a_2+a_3s)(a_6+a_7s)\]
Then, the functions $f,g$ and the discriminant $\Delta$ become
\ba
f(z)= -\lambda a_5a_6\,z^3,\; g(z)\sim z^5,\; \Delta \sim z^9
\label{E6up}
\ea
which upgrades the symmetry to $E_7\supset E_6\times U(1)$ whose fundamental decomposes
\[ 56\ra 27+\ov{27}+1+1\]

$b$) The case $a_5 a_6+ a_4 a_7=0$. This simultaneously   implies  $a_2=0$, as  can be observed
from the inspection of the Ansatz~(\ref{Ansatz}).
Together with  the previous condition $a_1=0$ implies
\[(a_4+a_5s)(a_6+a_7s) a_3 s^2=0\]
If we solve $a_5 a_6+ a_4 a_7=0$ assuming $a_5=\lambda a_4$, $a_7=-\lambda a_6$,
\[a_3 a_4a_6(1+\lambda s)(1-\lambda s) s^2=0\]
which implies ($a_3s^2=0$) that both $U(1)$'s involved in the monodromy descend
from $E_7$:
\[E_7\supset E_6\times U(1)\supset SO(10)\times U(1)\times U(1)'\]

$\bullet $ The $E_6$ enhancement of course occurs also if  $a_4=0$ (and similarly
if $a_6=0$). Notice that the ${\cal Z}_2$ monodromy is among the $t_{1,2}$ which are associated
to the coefficients $a_{1,2,3}$ and thus it  is now unaffected. This means that
the $U(1)$'s associated to $a_4$ (or $a_6$) coefficients are incorporated into the
$E_6$ symmetry. Therefore the monodromy  occurs among the $U(1)$'s emerging from
the commutant of  $E_6$:
\[E_8\ra E_6\times SU(3)\]
We proceed with the discriminant and $f,g$  functions which for $a_4=0$
are given by
\ba
f(z)&=&\left( a_1 a_5 a_7- \lambda a_5^2 a_6^2\right)z^3+\cdots,\nn
\\
g(z)&=&\frac{1}{4} a_1^2 a_5^2 a_6^2 z^4+\cdots,\nn
\\
\Delta(z)&=&-27 a_1^4 a_5^4 a_6^4\,z^8+\cdots\nn
\ea

$a$)
Let  now $a_1=0$. The case reduces to the previous one of $E_7$ enhancement an in (\ref{E6up}).

$b$) Let $a_6=0$.   Then
\[ f(z)= a_1 a_5 a_7z^3+\cdots,\; g(z)=\lambda a_5^2 a_7^2\,z^5+\cdots,\;\Delta(z)=-64 a_1^3 a_5^3 a_7^3z^9+\cdots\]
This is again an $E_7$ enhancement. The spectral cover equation is
\[(a_1+a_2s+a_3s^2)a_5a_7 s^2=0\]

{\bf $SO(12)$ enhancement}

The $SO(12)$ is obtained setting $b_3=0$.  This has two solutions: either
\[ a_5 a_6+a_4 a_7=0\Rightarrow \{ a_5\to \lambda a_4, a_7\to -\lambda a_6\}\]
or
\[a_1\to \lambda a_4 a_6\]

Substituting the first one in $b_k(a_i)$'s  of (\ref{so12en}) we get
\[\left(\lambda a_4 a_7+a_5 \left(\lambda a_6-\left(a_1+\lambda a_4 a_6\right) a_7\right)\right){}^2-4 \lambda a_1 a_4 a_5^2 a_6   a_7^2=k^4 a_4^2 a_6^2 \left(a_1-\lambda a_4 a_6\right){}^2\]
  If both conditions implying $b_3=0$ are imposed,  the latter is also zero, leading
  to $SO(14)$ enhancement. This enhancement can also  happen if instead of
    the second condition  $a_1\to \lambda a_4 a_6$ we impose $a_4=0$ or $a_6=0$.
  \begin{table}
\begin{center}
\begin{tabular}{|l|l|l|}
\hline
$E_6$&$E_7$&Monodromy \\
\hline
$a_1=0$&$a_4/a_6=0$&$ t_{1,2}\in E_6\times SU(3) $
\\
$a_1=0$&$a_5a_6+a_4a_7=0$&$t_{1,2}\in E_7$
\\
$a_4=0$&$a_6=0$&$\tilde t_{1,2}\in SU(3)$
\\$a_6=0$&$a_4=0$&$\tilde t_{1,2}\in SU(3)$
\\
\hline
\end{tabular}
\caption{The vanishing coefficients with the corresponding enhancements and the embedding of the
$U(1)$'s involved in the monodromy. In the first case the monodromy is between a $U(1)\in E_6$
and $U(1)\in SU(3)$. In the last two cases, the monodromy is among the $U(1)$'s in the
orthogonal complement of $E_6$ implying identification of the $27$'s. }
\label{T_a}
\end{center}
\end{table}
 We summarize all the above cases in Table~\ref{T_a}. In the last two entries of this Table
 the $\tilde t_{1,2}$ are incorporated into the spectral cover.

\subsubsection{$SU(5)$ and spectral cover splitting for the ${\cal Z}_2$ case}

 The case of  ${\cal Z}_2$  monodromy corresponds to the
following splitting of the spectral cover equation
\ba
b_0\prod_{i}(s-t_i)=(a_1+a_2s+a_3s^2)(a_4+a_5s)(a_6+a_7s)(a_8+a_9s)=\sum_{k=0}^5 b_k s^{5-k}\label{5split}
\ea
Following~\cite{Dudas:2010zb} we derive the relations of $b_k(a_i)$ by
equating coefficients of the same powers in $s$
\be
\label{bksZ2}
\begin{split}
b_0&=a_3 a_5 a_7 a_9\\
b_1&=a_3 a_5 a_7 a_8+a_3 a_5 a_6 a_9+a_3 a_4 a_7 a_9+a_2 a_5 a_7 a_9\\
b_2&=a_3 a_5 a_6 a_8+a_3 a_4 a_7 a_8+a_2 a_5 a_7 a_8+a_3 a_4 a_6 a_9+a_2 a_5 a_6 a_9+a_2 a_4 a_7 a_9+a_1
   a_5 a_7 a_9\\
b_3&=a_3 a_4 a_6 a_8+a_2 a_5 a_6 a_8+a_2 a_4 a_7 a_8+a_1 a_5 a_7 a_8+a_2 a_4 a_6 a_9+a_1 a_5 a_6 a_9+a_1
   a_4 a_7 a_9\\
b_4&=a_2 a_4 a_6 a_8+a_1 a_5 a_6 a_8+a_1 a_4 a_7 a_8+a_1 a_4 a_6 a_9\\
b_5&=a_1 a_4 a_6 a_8
\end{split}
\ee
while to solve the constraint
\[0=b_1=  a_3 a_5 a_7 a_8+a_3 a_4 a_9 a_7+a_2 a_5 a_7 a_9 +a_3 a_5 a_6 a_9\]
we use the Ansatz
\[ a_2 =-c ( a_5 a_7 a_8+a_4 a_9 a_7+ a_5 a_6 a_9),\; a_3=c  a_5 a_7 a_9\]
Notice that setting $a_8=0, a_9=1$ this complies with our previous $SO(10)$ Ansatz
(\ref{Ansatz}).
\begin{table}
\begin{center}
\begin{tabular}{c|c|c|c|c|c|c|c|c}
\hline
$a_1$& $a_2$& $a_3$& $a_4$& $a_5$& $a_6$& $a_7$&$a_8$&$a_9$
\\
$\eta-2c_1-\chi$& $\eta-c_1-\chi$& $\eta-\chi$& $-c_1+x_5$& $x_5$& $-c_1+x_7$& $x_7$&$-c_1+\chi_9$&$\chi_9$\\
\hline
\end{tabular}
\end{center}
\caption{Homology classes for coefficients $a_i$  for the ${\cal Z}_2$ ($SU(5)$) case}
\label{T_0}
\end{table}
The tenplets are found by studying the zeroth order of the above polynomial, which is
\[ b_5=t_1t_2t_3t_4t_5=a_1a_4a_6a_8\]
These are designated as
 \ba
 10^{(1)}_{t_1},10^{(2)}_{t_3},10^{(3)}_{t_4},10^{(4)}_{t_5}
 \label{10sZ2X}
 \ea
while their homologies are associated to those of $a_1, a_4, a_6, a_8$.
To determine the properties of the fiveplets we need the corresponding spectral cover equation.
This is a 10-degree polynomial
\[ {\cal P}_{10}(s)\propto\sum_{n=1}^{10}c_ns^{10-n}=b_0\prod_{i,j}(s-t_i-t_j),\; i<j,\; i,j=1,\dots, 5\]
Using~(\ref{5split}) we can convert the coefficients $c_n=c_n(t_j)$  to functions of
$c_n(b_j)$.  In particular we are interested for the the value ${\cal P}_{10}(0)$
given by the coefficient $c_{10}$ which can be expressed in terms of $b_k$ according to
\ba
c_{10}(b_k)&=& b_3^2b_4-b_2b_3b_5+b_0b_5^2=0\label{c10_A}
\ea
Using the equations $b_k(a_i)$ and the Ansatz, we can split this equation into seven factors
which correspond to the seven distinct fiveplets left after the ${\cal Z}_2$ monodromy action
given in (\ref{5spitZ2}) and in Table~\ref{SU5Z2spec}.

\subsection{${\cal Z}_2\times {\cal Z}_2$ spectral cover factorization }

In the case of $C_{2+2}$ factorization we write the polynomial $P_{16}(s)$ as follows
\[P_{16}(s)=  \left(a_3 s^2+a_2 s+a_1\right) \left(a_6 s^2+a_5 s+a_4\right)\]
This splitting implies the $t_1\lra t_2$ and $t_3\lra t_4$ identifications.
Comparing with the coefficients $b_k$ we have
\be
\label{Z2Z2}
\begin{split}
b_4&=a_1 a_4\\
b_3&=a_2 a_4+a_1 a_5\\
b_2&=a_3 a_4+a_2 a_5+a_1 a_6\\
b_1&=a_3 a_5+a_2 a_6\\
b_0&=a_3 a_6
\end{split}
\ee
Starting from the equation $b_1=0$, we see that the acceptable Ansatz is
\be
\label{Ansatz2z2}
\begin{split}
 a_3&=\lambda\,a_6,\;\; a_2=-\lambda\,a_5
 \end{split}
 \ee
In this case the $b_k$'s are given by
\be
\label{Z2Z2S}
\begin{split}
b_4&=a_1 a_4\\
b_3&=\left(a_1-\lambda  a_4\right) a_5\\
b_2&=\left(a_1+\lambda  a_4\right) a_6-\lambda a_5^2\\
b_1&=0\\
b_0&=\lambda a_6^2
\end{split}
\ee
Repeating the same steps as in the ${\cal Z}_2$ case we determine the homologies of $a_i$ given in Table
\ref{T_22} and the properties of the matter curves given in Table \ref{Z22_ALL}.
\begin{table}
\begin{center}
\begin{tabular}{c|c|c|c|c|c}
\hline
$a_1$& $a_2$& $a_3$& $a_4$& $a_5$& $a_6$
\\
$\eta-2c_1-\chi$& $\eta-c_1-\chi$& $\eta-\chi$& $\chi-2c_1$& $\chi-c_1$& $\chi$\\
\hline
\end{tabular}
\end{center}
\caption{Homology classes for coefficients $a_i$ for the ${\cal Z}_2\times {\cal Z}_2$  case.}
\label{T_22}
\end{table}
Next we recall that the $E_6$  enhancement is obtained for $b_4=a_1a_4=0$ and this can be done
by demanding either of $a_{1,4}$ to be zero.  The case $a_1=0$  implies
\[{\cal C}_4= s (a_2+a_3 s)(a_4+a_5s+a_6s^2) \]
Thus for the particular choice $a_1=0$ at the first order enhancement
to $E_6$ one ${\cal Z}_2$ monodromy is `resolved' with one of the two $U(1)$
incorporated into the $E_6$ symmetry. (The same happens for $a_4=0$.)
The $E_7$ enhancement is reached when in addition we have  $b_3=0$ and this
happens   either when $a_1=a_4=0$ or $a_1=a_5=0$ (or $a_4=0=a_5=0$).
The cases are collected in Table~\ref{T_bb}.

  \begin{table}
\begin{center}
\begin{tabular}{|l|l|l|}
\hline
$E_6$&$E_7$& $U(1)$ embedding $i=1,2,\,j=3,4$ \\
\hline
$a_1=0$&$a_4=0$&$ t_{i,j}\in E_7\times SU(2) $
\\
$a_1=0$&$a_5=0$&$t_{i}\in E_7,\;\tilde t_{j}\in SU(2)$
\\
$a_4=0$&$a_5=0$&$\tilde t_{i}\in SU(2),\;t_{j}\in E_7$
\\
\hline
\end{tabular}
\caption{The vanishing coefficients with the corresponding enhancements and the embedding of the
$U(1)$'s involved in the monodromy. In the first case the monodromies are between a $U(1)\in E_7$
and a $U(1)\in SU(2)$.}
\label{T_bb}
\end{center}
\end{table}

\subsubsection{The $SU(5)$ case}

In the ${\cal Z}_2\times {\cal Z}_2$ case, the spectral cover equation for $SU(5)$ obtains the form
\[{\cal C}_{10}(s)=  \left(a_3 s^2+a_2 s+a_1\right) \left(a_6 s^2+a_5 s+a_4\right) (a_7+a_8 s)\]
Proceeding as in the $SO(10)$ case, we identify the relations
$b_k(a_i), k=1,\dots 5$ by comparing coefficients of the same power in $s$.
\begin{equation}
\begin{split}
b_0&=a_3 a_6 a_8\\
b_1&=a_3 a_6 a_7+a_3 a_5 a_8+a_2 a_6 a_8\\
b_2&=a_3 a_5 a_7+a_2 a_6 a_7+a_3 a_4 a_8+a_2 a_5 a_8+a_1 a_6 a_8\\
b_3&=a_3 a_4 a_7+a_2 a_5 a_7+a_1 a_6 a_7+a_2 a_4 a_8+a_1 a_5 a_8\\
b_4&=a_2 a_4 a_7+a_1 a_5 a_7+a_1 a_4 a_8\\
b_5&=a_1 a_4 a_7
\end{split}
\end{equation}
The constraint  $b_1=0$ is solved by the Ansatz
 \be
 \label{Ansatz2z25}
 \begin{split}
  a_3&=\lambda\, a_6a_8,\; a_2=-\lambda ( a_6 a_7 + a_5 a_8 )
  \end{split}
  \ee
  Notice that this reduces to $SU(4)$ case (\ref{Ansatz2z2}) when $a_7=0, a_8=1$.
  Solving the equations analogously to the previous cases, we can easily determine the homology
  classes of $a_i$ given in Table~\ref{22HC}.
    \begin{table}
\begin{center}
\begin{tabular}{c|c|c|c|c|c|c|c}
\hline
$a_1$& $a_2$& $a_3$& $a_4$& $a_5$& $a_6$& $a_7$& $a_8$
\\
$\eta-2c_1-\chi-\psi$& $\eta-c_1-\chi-\psi$& $\eta-\chi-\psi$& $\chi-2c_1$& $\chi-c_1$& $\chi$ &$-c_1+\psi$&$\psi$\\
\hline
\end{tabular}
\end{center}
\caption{Homology classes for coefficients $a_i$ for the ${\cal Z}_2\times {\cal Z}_2$ case for $SU(5)$.}
\label{22HC}
\end{table}

The $10\in SU(5)$ are obtained from the solutions of the equation
\[ b_5=0,\; \ra a_1a_4a_7=0\]
therefore they are associated to $a_1=0, a_4=0$ and $a_7=0$.
The fiveplets are found  by solving  the corresponding equation
\[ b_3^2b_4-b_2b_3b_5+b_0b_5^2=0\]
  It is straightforward to determine the homology classes and other properties
  using the results of table \ref{22HC}.   The results are summarized
  in \ref{P5SU5} and in Table \ref{SU5Z2Z2} in the main body of the paper.

\subsection{${\cal Z}_3$ factorization}

We write the spectral cover equation as follows
\[{\cal C}_4=\left(a_1+a_2 s+a_3 s^2+a_4 s^3\right) \left(a_5+s a_6\right)\]
Comparing with the coefficients $b_k$, we get
\be
\label{Z3}
\begin{split}
b_4&=a_1 a_5\\
b_3&=a_2 a_5+a_1 a_6\\
b_2&=a_3 a_5+a_2 a_6\\
b_1&=a_4 a_5+a_3 a_6\\
b_0&=a_4 a_6
\end{split}
\ee
Imposing the conditions $a_5=\lambda a_6,\, a_3=-\lambda a_4$ we get
\be
\label{Z3S}
\begin{split}
b_4&=\lambda a_1 a_6\\
b_3&=(a_1+\lambda a_2) a_6\\
b_2&=(a_2-\lambda^2 a_4) a_6\\
b_1&=0\\
b_0&=a_4 a_6
\end{split}
\ee
The $b_4=0$ enhancement to $E_6$ is obtained by $a_1=0$ while  the
condition $b_3=0$ associated to
 the $E_7$ enhancement is obtained by setting $a_2=0$. Notice however
 that in this Ansatz the condition $a_6=0$ eliminates all
$b_k$'s.
A non-trivial solution is given by the Ansatz  $a_4=\lambda a_6,\, a_3=-\lambda a_4$
which entails the following forms of $b_k$'s
\be
\label{Z3SA}
\begin{split}
b_4&= a_1 a_5\\
b_3&=a_1a_6+a_2a_5\\
b_2&=a_2 a_6-\lambda\, a_5^2\\
b_1&=0\\
b_0&=\lambda\, a_6^2
\end{split}
\ee
It is straightforward to correlate the $E_6,E_7$ enhancements with the vanishing of
the appropriate coefficients $a_i$.

The homology classes of $a_i$ are shown in Table \ref{T_Z3}.
\begin{table}[t] \centering%
\begin{center}
\begin{tabular}{c|c|c|c|c|c}
\hline
$a_1$& $a_2$& $a_3$& $a_4$& $a_5$& $a_6$
\\
$\eta-3c_1-\chi$& $\eta-2c_1-\chi$& $\eta-c_1-\chi$& $\eta-\chi$& $\chi-c_1$& $\chi$\\
\hline
\end{tabular}
\end{center}
\caption{Homology classes for coefficients $a_i$ for the ${\cal Z}_3$ case in the $SO(10)$  model.}
\label{T_Z3}
\end{table}
The results are collected in Table~\ref{T16Z3}.

\subsubsection{ $SU(5)$  with ${\cal Z}_3$ monodromy}

In this case the relevant spectral cover polynomial  is
\[\sum_{k=0}^5 b_ks^{5-k}=\left(a_4 s^3+a_3 s^2+a_2 s+a_1\right) \left(a_5+s a_6\right) \left(a_7+s a_8\right)\]
We can easily extract the equations  determining the coefficients
$b_k(a_i)$, by equating equal powers of $s$
\be
\label{Z3bs}
\begin{split}
b_0&=a_4 a_6 a_8\\
b_1&=a_4 a_6 a_7+a_4 a_5 a_8+a_3 a_6 a_8\\
b_2&=a_4 a_5 a_7+a_3 a_6 a_7+a_3 a_5 a_8+a_2 a_6 a_8\\
b_3&=a_3 a_5 a_7+a_2 a_6 a_7+a_2 a_5 a_8+a_1 a_6 a_8\\
b_4&=a_2 a_5 a_7+a_1 a_6 a_7+a_1 a_5 a_8\\
b_5&=a_1 a_5 a_7
\end{split}
\ee
For each of the above equations, there is a corresponding one  for the homologies
\[[b_k]=\eta-k c_1= [a_l]+[a_m]+[a_n],\;k=0,1,\dots,5,\; k+l+m+n=18, \;l,m,n\le 8\]
the latter being valid only for the  combinations of indices appearing in (\ref{Z3bs}).
The homology classes of $a_i$ are shown in Table \ref{T5_Z3}.
\begin{table}[t] \centering%
\begin{center}
\begin{tabular}{c|c|c|c|c|c|c|c}
\hline
$a_1$& $a_2$& $a_3$& $a_4$& $a_5$& $a_6$&$a_7$&$a_8$
\\
$\eta-3c_1-\chi-\psi$& $\eta-2c_1-\chi-\psi$& $\eta-c_1-\chi-\psi$& $\eta-\chi-\psi$& $\chi-c_1$& $\chi$&$\psi-c_1$&$\psi$\\
\hline
\end{tabular}
\end{center}
\caption{$SU(5)$: Homology classes for coefficients $a_i$ for the ${\cal Z}_3$ case.}
\label{T5_Z3}
\end{table}
 The condition
\[0=b_1=a_4 a_6 a_7+a_4 a_5 a_8+a_3 a_6 a_8\]
is solved using the Ansatz
\[a_3\to -c \left(a_6 a_7+a_5 a_8\right), a_4\to c a_6 a_8\]
which, again as expected,  reduces to the corresponding $SU(4)$ case
(\ref{Z3SA}) when $a_7=0, a_8=1$.
Substitution of this conditions into the solution $b_k(a_i)$ gives
\be
\begin{split}
b_0&=c a_6^2 a_8^2\\
b_1&=0\\
b_2&=\left(a_2-c a_5 a_7\right)a_6 a_8 -c (a_6^2 a_7^2+ a_5^2 a_8^2)\\
b_3&=(a_6 a_7 +a_5 a_8) \left(a_2-c a_5 a_7\right)+a_1 a_6 a_8\\
b_4&=(a_6 a_7+ a_5 a_8) a_1 +a_2 a_5 a_7\\
b_5&=a_1 a_5 a_7
\end{split}
\ee
The tenplets are determined by $b_5=a_1 a_5 a_7=0$.
Under the above Ansatz, in the case of ${\cal Z}_3$ monodromy
the equation  (\ref{c10}) factorizes as in (\ref{Z35split}).
 These, together with the tenplets are given in Table~\ref{SU5Z3spec}.

\end{document}